# Multi-Unit Facility Location Games*

Omer Ben-Porat and Moshe Tennenholtz
Technion - Israel Institute of Technology
{omerbp@campus, moshet@ie}.technion.ac.il

Facility location games have been a topic of major interest in economics, operations research and computer science, starting from the seminal work by Hotelling. In the classical pure location Hotelling game businesses compete for maximizing customer attraction by strategically locating their facilities, assuming no price competition, while customers are attracted to their closest facilities. Surprisingly, very little rigorous work has been presented on multi-unit facility location games, where the classical pure location Hotelling games are extended to contexts where each player is to locate several facilities. In this paper we present two major contributions to the study of multi-unit pure location Hotelling games. In the first part of this paper we deal with the two-player multi-unit setting, and fully characterize its equilibria. In the second part of this paper we deal with multi-unit facility location games, with $N \geq 3$ players; our main result in this part is the full characterization of the settings where pure strategy equilibria exist. Our results also extend classical results on necessary and sufficient conditions for a strategy profile to be in equilibrium in a pure location (single-unit) Hotelling game to multi-unit pure location games.

**1. Introduction** In his seminal work [9], Hotelling introduced a canonical model of competition among businesses. In a classical motivating scenario Hotelling considered two ice-cream vendors, who sell ice-cream to sunbathers on the beach, and wish to maximize their payoffs. The vendors sell the same type of product, and charge the same price. Sunbathers are distributed uniformly along the beach and every sunbather walks to his/her nearest ice-cream vendor to buy an ice-cream. As indicated by Hotelling, the vendors will strategically locate their ice-cream carts in the middle of the beach, back to back, as this is the only Nash equilibrium of this game. Following that seminal work, facility location games have been a topic of major interest in economics, operations research and computer science.

A natural extension of this location game is when players operate, and so must locate, several facilities. In the above example, this corresponds to each ice-cream vendor running several stands along the shore. Similar situations occur when coffee shop chains have to decide where to locate stores in a newly created shopping strip in a big city. Surprisingly, very little rigorous work has been published on such *multi-unit* facility location games although some exceptions are discussed below.

In this paper we consider the multi-unit version of the pure location Hotelling game. The term pure location is used to emphasize that the products and the prices offered by all players are identical, with the only difference relevant to the customer being their distances from the corresponding facilities. In other words it assumes that players' payoffs and customers' choice are determined solely by the location of the facilities and the customers. It is further assumed that customers will visit the nearest facility.

Our model is mathematically formalized in Section 2, and can be briefly described as follows: we consider a set of customers distributed on the $[0,1]$ interval, and $N$ players, where player $i$ can locate

---

$n_i$ facilities on that interval. A customer is attracted to the nearest facility. The payoff of a player is the expected number of customers attracted to her facilities. We consider a uniform distribution of customers as in the original Hotelling setting and most extensions of pure location Hotelling games. The case where $n_i = 1$ for every $i$ corresponds to the classical pure location Hotelling game. Notice that due to the infinite action space, the existence of equilibrium is not immediate.

In Sections 3-4 we consider $N = 2$, the two-player game. We let player 1 locate $l$ facilities and player 2 locate $k$ facilities, where $l \leq k$. We examine the optimization problem where a (monopoly) player has to locate $k$ facilities in a way that will minimize the expected absolute distance of customers from their nearest facilities (called the social cost hereinafter), and highlight the importance of these $k$-socially optimal locations. We show that if player 2 locates her facilities in the $k$-socially optimal locations, and player 1 selects an $l$-tuple uniformly at random from these locations, equilibrium is obtained. After proving the corresponding existence result in Section 3, in Section 4 the equilibria are completely characterized and shown to all have the same structure. In particular, our results imply that in any (possibly mixed) equilibrium, facilities are located in the $k$-socially optimal locations only. Notice that this result is obtained although in all cases where the players control a different number of facilities there is no pure strategy equilibrium.

In Section 5 we deal with the case of $N \geq 3$. We generalize the known necessary and sufficient conditions for a pure strategy profile to be in equilibrium (see [5]) in the classical (i.e. single-unit) pure location Hotelling games. We then prove that a multi-unit facility location game possesses a pure strategy equilibrium if and only if there is no dominant player – a player who controls more than half of the facilities. We show that when a dominant player is present, there are conditions under which a mixed equilibrium exist; the facilities are located in the positions that would have been selected by the dominant player if she were a monopoly minimizing the social cost, generalizing our finding for the two-player game.

Beyond their classical economic appeal, multi-unit facility location games are applicable to recent data science contexts. This suggests a new motivation to the classical Hotelling setting and its extension to the multi-unit scenario. These future directions will be discussed in Section 6.

**1.1. Related Work** An informative survey of facility location games literature up to 1992 appears in [2]. For a more recent survey the reader is referred to [4]. Many extensions of the above basic Hotelling setting have been considered, although in most of the existing literature the set of possible actions (i.e. possible facility locations) is very simple, e.g, a segment or a circle. For a recent discussion of general graph structures in this setting see [8]. Within the literature related to pure location Hotelling games little consideration has been given to players controlling several facilities that need to be located in their actions. There are some exceptions, as we describe next.

The work of [17] considers the pure location competition between two players, who control more than one facility, to be located on the $[0, 1]$ segment with uniform customer distribution. This is the setting we consider when restricting the game to two players. While the paper does not discuss the pure location game equilibrium, the author presents a minimax strategy for the dominant player (who owns more facilities) while neglecting the analysis of the non-dominant player and the overall simultaneous game. Notice that the existence of a minimax strategy for a player does not imply the existence of equilibrium in infinite games. Indeed, our work is the first to show the existence of equilibrium for this setting.

The work of [10] considers pure location Hotelling games, where a coalition of players may collude. One may equate a colluding coalition to a player who owns several facilities. While most of that work does not deal with equilibrium analysis, one of the results there [10, Proposition 6] can be interpreted as showing mixed equilibria for some of the instances discussed in Lemma 11 of this work. Note that most of our study on multiple player games is devoted to full characterization of the situations where pure strategy equilibria exist.

Pure location Hotelling games are closely related to Voronoi games (see e.g. [1, 6, 12]). In a Voronoi game, players alternatingly place points in some space, and then every player gains the total surface of the Voronoi cells of his points. In [1] the authors consider a two-player symmetric game on the segment, where the players locate facilities alternatively, and are not able to locate a facility on an already occupied location. Our model is different, since a) we consider a simultaneous game, b) we analyze the case of any number of players and c) players can locate facilities in the same locations. In [1] the authors search for a sub-game perfect equilibria, whereas in our model the game is simultaneous, and mixed strategy equilibria are the solution concept to be analyzed.

Few other works consider particular multi-stage facility location games with multiple units, in settings vastly different from pure location Hotelling games and from the above extensions. In [14] the authors study the sub-game perfect equilibria of a particular two-stage facility location, where in stage one the firms choose the locations of their facilities, and in stage two the firms compete in quantities as in a Cournot competition. It is interesting to note that in their setup $n$ players with $m$ facilities each, will all locate their facilities in the monopoly socially-optimal locations. Hence, there is a vast difference between their setup to pure location Hotelling games and our results, where three or more facilities cannot be agglomerated in the same location in equilibrium. In [11] the authors consider an elaborated two-firm three-stage model of facility location on a circle. The firms select the number of facilities in the first stage, their locations in the second stage, and their prices in the third stage. It is interesting to note that in their setup the locations selected by the players in the second stage, in a sub-game perfect equilibrium, depend only on the number of selected facilities. This is in sharp contrast to the results of our work, where the locations selected by the non-dominant player depends on the number of facilities available to the dominant player.

**2. Model and Preliminaries** The non-cooperative game we consider is formally defined as follows:

1. A continuous density function $g$ over the unit interval $[0,1]$, representing customer distribution.
2. A set of players $[N] = \{1, 2, \ldots, N\}$, where the (pure) strategy set of player $i \in [N]$ is $S_i = \{(s_i^1, s_i^2, \ldots, s_i^{n_i}) \in [0,1]^{n_i} : s_i^1 < s_i^2 < \cdots < s_i^{n_i}\}$. $(n_i)_{i \in [N]}$ are exogenously determined, and w.l.o.g. we assume $n_1 \leq n_2 \leq \cdots \leq n_N$. It will sometimes be convenient to say that player $i$ has/owns a *facility* in $s_i^j$ under $s_i$. A mixed strategy of every player $i \in [N]$ is an element of $\Delta(S_i)$, which is the set of all probability distributions over $(S_i, \mathcal{B}(S_i))$, where $\mathcal{B}(S_i)$ is the standard Borel $\sigma$-algebra over $S_i$.
3. Given a pure strategy profile $\mathbf{s} = (s_1, \ldots, s_N) \in \prod_{i=1}^N S_i$ and $t \in [0,1]$, let

$$r_t(\mathbf{s}) = \left\{ i \in [N] : \min_j |t - s_i^j| \leq \min_{i',j'} |t - s_{i'}^{j'}| \right\}.$$

Namely, $r_t(\mathbf{s})$ is the set of players whose facilities are the closest to the customer $t \in [0,1]$. As is standard, we assume that every customer is attracted to his nearest facility. The payoff of player $i$ under a pure strategy profile $\mathbf{s}$ is given by

$$u_i(\mathbf{s}) = \int_0^1 \frac{\mathbb{1}_{i \in r_t(\mathbf{s})}}{|r_t(\mathbf{s})|} g(t) dt,$$

and is extended to mixed strategy profiles $\mathbf{x} = (x_1, \ldots, x_N) \in \prod_i \Delta(S_i)$ by

$$u_i(\mathbf{x}) = \mathop{\mathbb{E}}_{\mathbf{s} \sim \mathbf{x}} u_i(\mathbf{s}).$$

In this paper, we focus on the uniform density function $g$, and refer to other cases in Section 6. According to these assumptions, every game is fully described by the number of players and the number of facilities of each player, i.e. $G = (N, (n_i)_{i \in N})$. [1]

**Equilibrium**

For a vector $\mathbf{v} = (v_1, \ldots v_n)$, we denote by $\mathbf{v}_{-i} := (v_1, \ldots v_{i-1}, v_{i+1}, \ldots v_n)$ the vector that does not contain the $i$-th coordinate of $\mathbf{v}$.

A strategy profile $\mathbf{s} \in \prod_{i \in [N]} S_i$ is called a *pure Nash equilibrium* if for every $i \in [N]$ and every $s'_i \in S_i$ it holds that $u_i(s'_i, \mathbf{s}_{-i}) \leq u_i(\mathbf{s})$.

Similarly, a mixed strategy profile $\mathbf{x} \in \prod_{i \in [N]} \Delta(S_i)$ is a *Nash equilibrium* if for every $i \in [N]$ and every $x'_i \in \Delta(S_i)$ it holds that $u_i(x'_i, \mathbf{x}_{-i}) \leq u_i(\mathbf{x})$.

We say that $x_i \in \Delta(S_i)$ is *a best response* to $\mathbf{x}_{-i} \in \prod_{j \in [N] \setminus \{i\}} \Delta(S_j)$ if $u_i(x_i, \mathbf{x}_{-i}) = \sup_{x'_i \in \Delta(S_i)} u_i(x'_i, \mathbf{x}_{-i})$. It is well known that if $x_i$ is a best response to $\mathbf{x}_{-i}$, there exists a pure strategy $s_i \in S_i$ such that $s_i$ is a best response to $\mathbf{x}_{-i}$.

We say that player $i$ has a *beneficial deviation* under a mixed strategy profile $\mathbf{x} = (x_i, \mathbf{x}_{-i})$ if there exists $x'_i \in \Delta(S_i)$ such that $u_i(x'_i, \mathbf{x}_{-i}) > u_i(\mathbf{x})$.

**Notations** There exists a bijection from strategies (strictly increasing vectors) to sets, $s_i \mapsto \cup_{j \in [n_i]} \{s_i^j\}$. Thus, for expositional simplicity, we refer to pure strategies as sets and as vectors interchangeably. For instance, given a strategy $s_i = (s_i^1, \ldots, s_i^{n_i})$, if there exists an index $j$ such that $f = s_i^j$, we denote $f \in s_i$. In addition, if $s_i = (f, s_i^2, \ldots, s_i^{n_i})$ and $s'_i = (f', s_i^2, \ldots, s_i^{n_i})$, it will be comfortable to denote $s'_i = s_i \setminus \{f\} \cup \{f'\}$.

Notice that a player's payoff is the total customer mass attracted to her facilities, or equivalently the sum of customer mass attracted to each of her facilities. In this sense, facilities, even those owned by the same player, are competing over customer mass. Accordingly, the customer mass attracted to a facility of a player is determined by the other facilities' locations only, and not by their owners' identities. To use this intuition in the upcoming analysis, we denote by

$$\mathcal{L}(\mathbf{s}) = \cup_{i \in [N]} s_i$$

the set of locations selected by the players under the profile $\mathbf{s}$. Given a pure strategy profile $\mathbf{s}$ and a facility $f \in s_i$, let $\mathcal{V}_i(f; \mathbf{s})$ denote the customer mass attracted to $f \in s_i$. Namely,

$$\mathcal{V}_i(f; \mathbf{s}) = \begin{cases} \int_0^1 \frac{\mathbb{1}_{|t-f| \leq d(t, \mathcal{L}(\mathbf{s}))}}{|r_t(\mathbf{s})|} dt & f \in s_i \\ 0 & otherwise \end{cases},$$

where $d : 2^{[0,1]} \to [0,1]$ is defined as $d(t, A) = \inf_{f \in A} |t - f|$. Thus, an alternative representation of player payoffs is $u_i(\mathbf{s}) = \sum_{f \in s_i} \mathcal{V}_i(f; \mathbf{s})$.

**Social cost** In [9], Hotelling examines the transportation cost of a customer: the distance he has to travel to reach his nearest facility. This notion is here referred to as the *social cost*.

DEFINITION 1 (SOCIAL COST). Given a finite set of locations $A \subset [0,1]$, the social cost is the sum of distances customers must travel to reach their nearest facility. Formally, $\text{SC} : 2^{[0,1]} \to \mathbb{R}_{\geq 0}$ is defined by

$$\text{SC}(A) = \int_0^1 d(t, A) dt.$$

---

[1] We define the tie breaker as uniform selection among players (whose facilities are the closest to a customer) and not as uniform selection among the facilities themselves. An additional modeling decision is forbidding a player to locate two (or more) of her facilities in the same location. In fact, these two modeling decisions are taken for expositional simplicity only: the four corresponding modeling options induce the same game after eliminating dominant strategies. Breaking ties is relevant when two (or more) facilities are located in the same location, as otherwise it concerns a zero-measure set of customers. If a player cannot locate two of her facilities in the same location, the two tie breakers are equivalent (up to a zero-measure set). If a player is allowed to locate two of her facilities in the same location, it is easy to verify that under both tie breakers such strategies are strictly dominated.

The following proposition is a known result in the literature (see, e.g., [8]):

PROPOSITION 1. *Let $o^k = \{o_1^k, \ldots, o_k^k\}$, where $o_i^k = \frac{2i-1}{2k}$ for $i \in [k]$. It holds that*

$$\mathrm{SC}(o^k) = \inf_{A \subset [0,1], |A|=k} \mathrm{SC}(A).$$

For completeness, we present the proof in appendix A.1. We refer to the set $o^k$ as the *k-socially optimal locations*. Observe that under the optimal locations, the customer mass served by each facility is equal to $\frac{1}{k}$.

Notice that the social cost is a set function, and is influenced only by the locations in $A \subset [0,1]$. Hence, one can extend the social cost to pure strategy profiles, by considering all locations selected under the corresponding strategy profile, i.e.

$$\mathrm{SC}(\mathbf{s}) = \int_0^1 d(t, \mathcal{L}(\mathbf{s})) \, dt.$$

Analogously, the social cost of a mixed strategy profile is defined by $\mathrm{SC}(\mathbf{x}) = \mathbb{E}_{\mathbf{s} \sim \mathbf{x}} \mathrm{SC}(\mathbf{s})$.

**3. The Existence of an Equilibrium in Two-Player Games** In this section we show the existence of an equilibrium profile for two-player games. Consider $G = (2, (l, k))$. That is, $N = 2$, player 1 has $l$ facilities ($n_1 = l$) and player 2 has $k$ facilities ($n_2 = k$), and w.l.o.g. let $l \leq k$. For brevity, we refer to two-player games as $l$ vs $k$ games.

**3.1. Symmetric games** We first analyze the special case of $l = k$, i.e. $k$ vs $k$ games. The results for this symmetric game will be later used in our analysis of the general (asymmetric) case. Observe that the $k$-socially optimal locations are available to both players as pure strategies. We claim that each player can guarantee herself $\frac{1}{2}$ by playing $o^k$.

PROPOSITION 2. *For every $s \in S_2$ it holds that $u_2(o^k, s) \leq \frac{1}{2}$.* [2]

PROOF. Fix an arbitrary strategy $s \in S_2$. For every $f \in s$ exactly one of the following holds:
1. $f \in [0, o_1^k) \cup (o_k^k, 1]$. In this case, $\mathcal{V}_2(f; o^k, s) < \frac{1}{2k}$.
2. $f \in (o_j^k, o_{j+1}^k)$ for some $1 \leq j < k$; hence $\mathcal{V}_2(f; o^k, s) \leq \frac{1}{2k}$.
3. $f = o_j^k$ for some $j \in [k]$. Since $\mathcal{V}_1(o_j^k; o^k, s) + \mathcal{V}_2(f; o^k, s) \leq \frac{1}{k}$ and $\mathcal{V}_1(o_j^k; o^k, s) = \mathcal{V}_2(f; o^k, s)$, it follows that $\mathcal{V}_2(f; o^k, s) \leq \frac{1}{2k}$.

Therefore,

$$u_2(o^k, s) = \sum_{f \in s} \mathcal{V}_2(f; o^k, s) \leq \sum_{f \in s} \frac{1}{2k} = \frac{1}{2}.$$

□

Lemma 1 below implies that there is a unique Nash equilibrium for $k$ vs $k$ games, which is obtained when both players play $o^k$. This is exploited later, when we consider asymmetric games.

LEMMA 1. *For every $s \in S_2$ such that $s \neq o^k$, $u_2(o^k, s) < \frac{1}{2}$.*

PROOF. Fix an arbitrary strategy $s \in S_2$ such that $s \neq o^k$, and assume by contradiction that $u_2(o^k, s) = \frac{1}{2}$. According to Proposition 2, we conclude that

$$\mathcal{V}_2(f; o^k, s) \leq \frac{1}{2k} \tag{1}$$

---

[2] A similar observation appears in [17]. However, the model of [17] prohibits the players to agglomerate facilities; hence modifications are needed.

for all $f \in s$. Moreover, in order to have $u_2(o^k, s) = \frac{1}{2}$, Equation (1) must hold in equality; hence $\mathcal{V}_2(f; o^k, s) = \frac{1}{2k}$ for any $f \in s$. Denote

$$I_j = \left[o_j^k, o_{j+1}^k\right], \ 1 \leq j < k,$$

and let $I_0 = [0, o_1^k), I_k = (o_k^k, 1]$.

*Step 1:* Observe that no player 2 facility $f \in s$ is located in $I_0$, since if $f \in I_0$, $\mathcal{V}_2(f; o^k, s) < \frac{1}{2k}$, and similarly for $f \in I_k$. Thus, all player 2's facilities are located in $\cup_{i=1}^{k-1} I_i$, i.e. $|\cup_{i=1}^{k-1} I_i \cap s| = k$.

*Step 2:* If $I_j$ contains exactly two facilities owned by player 2, i.e. $|I_j \cap s| = 2$, it follows that $I_j \cap s = \{o_j^k, o_{j+1}^k\}$. To see this, denote these facilities by $I_j \cap s = \{f_1, f_2\}$ where w.l.o.g. $f_1 < f_2$. Now:

- If $o_j^k < f_1 < f_2 < o_{j+1}^k$,

$$\mathcal{V}_2(f_1; o^k, s) = \frac{f_2 - o_j^k}{2} < \frac{o_{j+1}^k - o_j^k}{2} = \frac{1}{2k}.$$

- If $o_j^k = f_1 < f_2 < o_{j+1}^k$, we have three cases: if $j = 1$, $f_1 = o_1^k$, then

$$\mathcal{V}_2(f_1; o^k, s) = \frac{f_2 + o_1^k}{4} < \frac{o_2^k + o_1^k}{4} = \frac{1}{2k},$$

and similarly for $j = k$. Alternatively, if $1 < j < k$, we have

$$\mathcal{V}_2(f_1; o^k, s) = \frac{1}{2}\left(\frac{f_2 - o_{j-1}^k}{2}\right) < \frac{o_{j+1}^k - o_{j-1}^k}{4} = \frac{1}{2k}.$$

A similar argument applies for the case where $o_j^k < f_1 < f_2 = o_{j+1}^k$. Since $\mathcal{V}_2(f_1; o^k, s) = \frac{1}{2k}$, we conclude that $o_j^k = f_1 < f_2 = o_{j+1}^k$.

*Step 3:* For every $1 \leq j < k$ we have $|I_j \cap s| < 3$. This is true since according to the previous step every pair of facilities in $I_j \cap s$ should be located on different endpoints of $I_j$, which has only two endpoints.

*Step 4:* Let

$$\alpha_{opt} = o^k \cap s, \alpha_{empt} = o^k \setminus s, \alpha_{int} = s \setminus o^k.$$

Namely, $\alpha_{opt}$ are the $k$-socially optimal locations occupied with player 2's facilities under $s$; $\alpha_{empt}$ are the $k$-socially optimal locations free of player 2's facilities under $s$; and $\alpha_{int}$ are player 2's facilities which are located outside $o^k$ (interior points of $I_1, \ldots I_{k-1}$).

By definition, $|\alpha_{opt}| + |\alpha_{empt}| = |\alpha_{opt}| + |\alpha_{int}| = k$; therefore $|\alpha_{empt}| = |\alpha_{int}|$. Since $s \neq o^k$, it follows that $|\alpha_{opt}| < k$, and $|\alpha_{empt}| = |\alpha_{int}| > 0$. Due to Step 1, $\alpha_{int} \cap (I_0 \cup I_k) = \emptyset$. Due to Steps 2 and 3, every facility in $f \in (\alpha_{int} \cap I_j)$ is the only facility of player 2 in $I_j$. Moreover, $f \in (\alpha_{int} \cap I_j)$ implies that $o_j^k, o_{j+1}^k \in \alpha_{empt}$, i.e. the endpoints of $I_j$ are not occupied with player 2's facilities. Thus, $|\alpha_{int}| \geq |\alpha_{empt}| + 1$, which yields the desired contradiction. We conclude that $|\alpha_{int}| = |\alpha_{empt}| = 0$, and $s = o^k$. □

Notice that $u_2(o^k, o^k) = \frac{1}{2}$, since both players choose the same locations. Hence, any strategy profile $(s_1, s_2)$ where $s_1 \neq o^k$ or $s_2 \neq o^k$ cannot be in equilibrium. This claim is extended to mixed strategies as well: if $x_2 \in \Delta(S_2)$ assigns positive probability to strategies other than $o^k$, then $u_2(o^k, x_2) < \frac{1}{2}$, and by symmetry, the same argument holds for every $x_1 \in \Delta(S_1)$.

COROLLARY 1. *In a k vs k game, the unique equilibrium is obtained when both players play the k-socially optimal strategy.*

PROOF. If $(x_1, x_2) \in \Delta(S_1) \times \Delta(S_2)$ is an equilibrium profile, then $u_1(x_1, x_2) = u_2(x_1, x_2) = \frac{1}{2}$ since otherwise the player who gets less can deviate to $o^k$ and guarantee herself $\frac{1}{2}$. In addition, if $\mathbb{P}_{s_1 \sim x_1}(s_1 \neq o^k) > 0$, Lemma 1 implies that $\mathbb{P}_{s_1 \sim x_1}\left(u_2(s_1, o^k) > \frac{1}{2}\right) > 0$; hence $u_2(x_1, o^k) > \frac{1}{2}$. Therefore, player 1 chooses $o^k$ with probability 1 under $x_1$, and symmetric argument applies for player 2. □

**3.2. General two-player games** Here we show the existence of an equilibrium profile for the $l$ vs $k$ game. We denote $o^{l,k}$ for $l \leq k$ the probability distribution that assigns probability $1/\binom{k}{l}$ to every subset of $o^k$ of size $l$. Formally:

$$\mathbb{P}_{o^{l,k}}(s) = \begin{cases} 1/\binom{k}{l} & \text{if } s \subseteq o^k, |s| = l \\ 0 & \text{otherwise} \end{cases}.$$

Note that $o^{l,k} \in \Delta(S_1)$. We show that

THEOREM 1. $(o^{l,k}, o^k)$ *is a Nash equilibrium of $l$ vs $k$ games.*

Notice that $u_1(o^{l,k}, o^k) = \frac{l}{2k}$ and $u_2(o^{l,k}, o^k) = 1 - \frac{l}{2k}$. To prove Theorem 1, one must show that none of the players has a beneficial deviation. This will be obtained using two supporting lemmas, namely Lemmas 2 and 3.

LEMMA 2. *For every $s \in S_1$ it holds that $u_1(s, o^k) \leq \frac{l}{2k}$.*

The proof is similar to the proof of Proposition 2, and hence omitted. Since $u_1(o^{l,k}, o^k) = \frac{l}{2k}$, player 1 has no beneficial deviation.

In order to show that player 2 has no beneficial deviation, we exploit a connection between $k$ vs $k$ games and $l$ vs $k$ games, in Lemma 3. In fact, we prove a slightly stronger claim than the lack of beneficial deviation for player 2: we claim that if player 2 deviates from $o^k$, her payoff will strictly decrease.

LEMMA 3. *In an $l$ vs $k$ game, if $s_2 \neq o^k$, it holds that $u_2(o^{l,k}, s_2) < 1 - \frac{l}{2k}$.*

PROOF. Fix an arbitrary strategy $s_2 \in S_2$ of player 2 such that $s_2 \neq o^k$. We have

$$u_1(o^{l,k}, s_2) = \sum_{s_1 \in supp(o^{l,k})} u_1(s_1, s_2) \mathbb{P}_{o^{l,k}}(s_1) = \sum_{s_1 \in supp(o^{l,k})} \sum_{i=1}^{k} \left( \mathcal{V}_1(o_i^k; s_1, s_2) \mathbb{P}_{o^{l,k}}(s_1) \right). \quad (2)$$

Since for any $s_1 \in supp(o^{l,k})$ we know that $s_1 \subseteq o^k$, for every $f \in s_1$ it follows that

$$\mathcal{V}_1(f; s_1, s_2) \geq \mathcal{V}_1(f; o^k, s_2),$$

since the mass of customers for which the facility in $f$ is the closest can only decrease. We derive from Equation (2) that

$$u_1(o^{l,k}, s_2) \geq \sum_{s_1 \in supp(o^{l,k})} \sum_{i=1}^{k} \left( \mathcal{V}_1(o_i^k; o^k, s_2) \mathbb{P}_{o^{l,k}}(s_1) \mathbb{1}_{o_i^k \in s_1} \right) = \sum_{i=1}^{k} \mathcal{V}_1(o_i^k; o^k, s_2) \overbrace{\sum_{s_1 \in supp(o^{l,k})} \mathbb{P}_{o^{l,k}}(s_1) \mathbb{1}_{o_i^k \in s_1}}^{\frac{l}{k}},$$

where the last argument follows directly from the definition of $o^{l,k}$. It follows that

$$u_1(o^{l,k}, s_2) \geq \frac{l}{k} \sum_{i=1}^{k} \mathcal{V}_1(o_i^k; o^k, s_2) = \frac{l}{k} u_1(o^k, s_2).$$

Notice that $u_1(o^k, s_2)$ is the payoff of player 1 in the symmetric $k$ vs $k$ game under the strategy profile $(o^k, s_2)$. According to Lemma 1 it holds that $u_2(o^k, s_2) < \frac{1}{2}$, and since the game is fixed-sum, $u_1(o^k, s_2) > \frac{1}{2}$, we can say $u_1(o^{l,k}, s_2) > \frac{1}{2} \frac{l}{k} = \frac{l}{2k}$. Finally,

$$u_1(o^{l,k}, s_2) > \frac{l}{2k} \Rightarrow u_2(o^{l,k}, s_2) < 1 - \frac{l}{2k}.$$

□

Overall, $o^k$ is the only best response of player 2 against $o^{l,k}$.

PROOF OF THEOREM 1. Lemmas 2 and 3 imply that any change in player $i$'s strategy will not improve her payoff; hence $(o^{l,k}, o^k)$ is in equilibrium.

**4. The Uniqueness of Equilibrium in Two-Player Games** Here we show that under any equilibrium of $l$ vs $k$ games, the realizable set of locations is the $k$-socially optimal locations with probability 1. Namely, any equilibrium has the property that player 2 employs the strategy $o^k$, while player 1 uses some distribution over $o^k$. This implies that player 1 does not contribute in any equilibrium any new location, beyond what player 2 contributes, which are the locations that correspond to the $k$-socially optimal ones.

When we refer to pure strategies or to "simple" mixed strategies, the players' payoffs are easy to compute. However, when one wishes to analyze an arbitrary mixed strategy profile, a few obstacles arise. In order to address this, we employ a new point of view for the analysis.

DEFINITION 2. Given a mixed strategy $x \in \Delta(S_1)$, let $\mu_x(A)$ denote the expected number of player 1's facilities in the set $A \subseteq [0,1]$. Formally,

$$\mu_x(A) = \mathop{\mathbb{E}}_{(s_1^1, s_1^2, \ldots, s_1^l) \sim x} \left( \sum_{i=1}^{l} \mathbb{1}_{s_1^i \in A} \right).$$

Observe that

PROPOSITION 3. *For every $x \in \Delta(S_1)$, the induced $\mu_x$ is a measure.*

The proof is given in appendix A.2. Note that $\mu_x$ is the expected number of player 1's facilities in the set $A$ under the strategy $x$, so for all $x \in \Delta(S_1)$ it follows that $\mu_x([0,1]) = l$. Next, we use $\mu$ to narrow the search for possible equilibrium strategies for player 1.

DEFINITION 3. We say that a strategy $x \in \Delta(S_1)$ is an $(l,k)$-*socially optimal imitation*, or $(l,k)$-SOI for brevity, if

$$\forall i \in [k]: \ \mu_x\left(\{o_i^k\}\right) = \frac{l}{k}.$$

Note that being an $(l,k)$-SOI strategy implies that $\mu_x(o^k) = l$; hence the support of any such strategy is a subset of the discrete set

$$\left\{(f_1, \ldots, f_l) : \forall i \in [l], \ f_i \in o^k\right\}.$$

In particular, the previously defined $o^{l,k}$ is $(l,k)$-SOI, and $o^k$ is $(k,k)$-SOI. For some values of $l$ and $k$ there may exist more than one $(l,k)$-SOI strategy. For instance,

$$x = \begin{cases} \left(\frac{1}{8}, \frac{3}{8}\right) & w.p. \frac{1}{2} \\ \left(\frac{5}{8}, \frac{7}{8}\right) & w.p. \frac{1}{2} \end{cases}, \ y = \begin{cases} \left(\frac{1}{8}, \frac{7}{8}\right) & w.p. \frac{1}{2} \\ \left(\frac{3}{8}, \frac{5}{8}\right) & w.p. \frac{1}{2} \end{cases}$$

are two $(2,4)$-SOI strategies.

Although $(l,k)$-SOI strategies are not unique, they are *quasi-unique*, i.e. the measure in which every location is selected will be identical in any $(l,k)$-SOI strategy. This fact will be useful later in proving the quasi-uniqueness of equilibrium.

Notice the proof of Lemma 3 did not use specific properties of $o^{l,k}$, apart from the fact that $\mu_{o^{l,k}}(\{o_i^k\}) = \frac{l}{k}$ for all $i \in [k]$. Therefore, the following is immediate from Lemma 3.

COROLLARY 2. *Let $x_1 \in \Delta(S_1)$ be an $(l,k)$-SOI strategy of player 1. If $s_2 \neq o^k$, it holds that $u_2(x_1, s_2) < 1 - \frac{l}{2k}$.*

We now turn to analyze potential equilibrium profiles. A few properties should be satisfied under any equilibrium profile for this setting.

LEMMA 4. *If $(x_1, x_2) \in \Delta(S_1) \times \Delta(S_2)$ is an equilibrium profile of an $l$ vs $k$ game, the following assertions must hold:*
1. $u_1(x_1, x_2) = \frac{l}{2k}, \ u_2(x_1, x_2) = 1 - \frac{l}{2k}$.

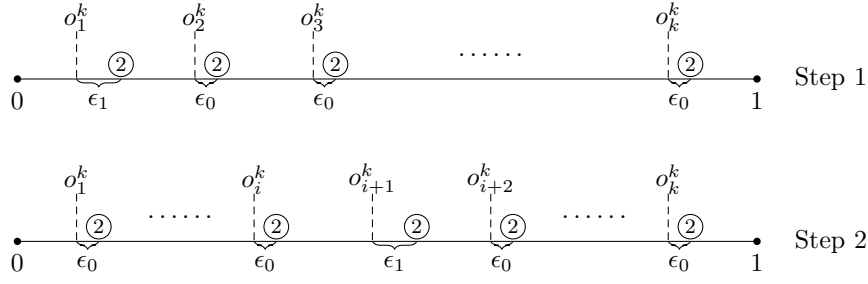

FIGURE 1. The beneficial deviations constructed for Steps 1 and 2. Each circle represents a facility of player 2. Consider the profile in Step 1, where we assume $\mu\left((o_1^k, o_2^k]\right) > \mu\left(o_1^k\right)$. The idea is to bracket facilities of player 1 inside the interval $(o_1^k + \epsilon_1, o_2^k + \epsilon_0)$, and make such facilities gain strictly less customer mass than under $s_2$, while potentially losing customer mass in the interval $(0, o_1^k + \epsilon_1)$. Notice that the distance between every two facilities $o_j^k, o_{j+1}^k$ for $1 < j < k$ remains as it was under $s_2$.

2. $\mu_{x_1}\left([0, o_1^k)\right) = \mu_{x_1}\left((o_k^k, 1]\right) = 0$.
3. $u_1(x_1, x_2) = u_1(x_1, o^k)$.

The proof of this lemma is deferred to appendix A.3. We are now ready to present the main result of this section.

THEOREM 2 (**Quasi-Uniqueness of Equilibrium**). *In an l vs k game, a strategy profile $(x_1, x_2)$ is in equilibrium if and only if*
**C1:** $x_1$ *is an $(l,k)$-SOI strategy, and*
**C2:** $x_2 = o^k$.

PROOF. If C1, C2 are satisfied, Corollary 2 and Lemma 2 imply that neither of the players has a beneficial deviation; hence $(x_1, x_2)$ is in equilibrium.

For the other direction: assume $(x_1, x_2)$ is an equilibrium strategy profile. In each of the following steps, we exploit the fact that $u_2(x_1, x_2) \geq u_2(x_1, s_2)$ for every $s_2 \in S_2$, or equivalently

$$u_1(x_1, s_2) \geq u_1(x_1, x_2) = \frac{l}{2k} \tag{3}$$

to obtain pieces of information about $x_1$.

To simplify the notations in the upcoming analysis, we remove the subscript $x_1$ from $\mu_{x_1}$, since $x_1$ is clear from the context. Moreover, we use $\mu(f)$ instead of $\mu(\{f\})$ for $f \in [0,1]$.

*Step 1*: We argue that $\mu\left(o_1^k\right) \geq \mu\left((o_1^k, o_2^k]\right)$. Assume by contradiction that $\mu\left((o_1^k, o_2^k]\right) - \mu\left(o_1^k\right) = \delta > 0$. Consider

$$s_2 = \left(o_1^k + \epsilon_1, o_2^k + \epsilon_0, \ldots, o_k^k + \epsilon_0\right)$$

for small enough constants $\epsilon_1, \epsilon_0 > 0$ such that[3] $\mu\left(o_i^k + \epsilon_j\right) = 0$ for all $i \in [k]$, $j \in \{0,1\}$ and

$$\mu\left((o_1^k, o_1^k + \epsilon_1)\right) < \frac{\delta}{6}, \quad \epsilon_0 < \frac{\delta \epsilon_1}{2l}. \tag{4}$$

Namely, $s_2$ is similar to $o^k$, but every facility is shifted right by $\epsilon_0$, except from the facility on $o_1^k$ which is shifted right by $\epsilon_1$. See Figure 1 for illustration.

---

[3] To justify the existence of such $\epsilon_0, \epsilon_1$, see Proposition 8 in the appendix. It should also be stated that $\epsilon_0, \epsilon_1 < \frac{1}{2k}$.

Notice that for every $s_1 \in S_1$, player 1's payoff satisfies

$$u_1(s_1, s_2) = \sum_{f \in s_1} \mathcal{V}_1(f; s_1, s_2)$$
$$= \sum_{f \in [0, o_1^k] \cap s_1} \mathcal{V}_1(f; s_1, s_2) + \sum_{f \in (o_1^k, o_1^k + \epsilon_1) \cap s_1} \mathcal{V}_1(f; s_1, s_2) + \sum_{f \in [o_1^k + \epsilon_1, o_2^k + \epsilon_0] \cap s_1} \mathcal{V}_1(f; s_1, s_2) + \sum_{f \in (o_2^k + \epsilon_0, 1] \cap s_1} \mathcal{V}_1(f; s_1, s_2)$$
$$\leq \sum_{f \in [0, o_1^k] \cap s_1} \left( \frac{1}{2k} + \frac{\epsilon_1}{2} \right) + \sum_{f \in (o_1^k, o_1^k + \epsilon_1) \cap s_1} \left( \frac{1}{2k} + \epsilon_1 \right) + \sum_{f \in [o_1^k + \epsilon_1, o_2^k + \epsilon_0] \cap s_1} \left( \frac{1}{2k} + \frac{\epsilon_0 - \epsilon_1}{2} \right) + \sum_{f \in (o_2^k + \epsilon_0, 1] \cap s_1} \frac{1}{2k}. \quad (5)$$

Taking expectation over Equation (5), we get

$$u_1(x_1, s_2) \leq \mu\left([0, o_1^k]\right) \left( \frac{1}{2k} + \frac{\epsilon_1}{2} \right) + \mu\left((o_1^k, o_1^k + \epsilon_1)\right) \left( \frac{1}{2k} + \epsilon_1 \right)$$
$$+ \mu\left([o_1^k + \epsilon_1, o_2^k + \epsilon_0]\right) \left( \frac{1}{2k} + \frac{\epsilon_0 - \epsilon_1}{2} \right) + \mu\left((o_2^k + \epsilon_0, 1]\right) \frac{1}{2k}.$$

By additivity of $\mu$, and due to $\mu([0, 1]) = l$, we have

$$u_1(x_1, s_2) \leq \frac{l}{2k} + \frac{\epsilon_1}{2} \mu\left([0, o_1^k]\right) + \epsilon_1 \mu\left((o_1^k, o_1^k + \epsilon_1)\right) + \frac{\epsilon_0 - \epsilon_1}{2} \mu\left([o_1^k + \epsilon_1, o_2^k + \epsilon_0]\right).$$

Lemma 4 guarantees that there are no facilities owned by player 1 in $[0, o_1^k)$ with probability 1, namely $\mu([0, o_1^k)) = 0$; thus

$$u_1(x_1, s_2) \leq \frac{l}{2k} + \frac{\epsilon_1}{2} \mu\left(o_1^k\right) + \epsilon_1 \mu\left((o_1^k, o_1^k + \epsilon_1)\right) + \frac{\epsilon_0 - \epsilon_1}{2} \mu\left([o_1^k + \epsilon_1, o_2^k + \epsilon_0]\right)$$
$$= \frac{l}{2k} - \frac{\epsilon_1}{2} \left( \mu\left([o_1^k + \epsilon_1, o_2^k + \epsilon_0]\right) - \mu\left(o_1^k\right) - 2\mu\left((o_1^k, o_1^k + \epsilon_1)\right) \right) + \frac{\epsilon_0}{2} \mu\left([o_1^k + \epsilon_1, o_2^k + \epsilon_0]\right).$$
$$\leq \frac{l}{2k} - \frac{\epsilon_1}{2} \left( \mu\left([o_1^k + \epsilon_1, o_2^k + \epsilon_0]\right) - \mu\left(o_1^k\right) - 2\mu\left((o_1^k, o_1^k + \epsilon_1)\right) \right) + \frac{l\epsilon_0}{2}, \quad (6)$$

where the last inequality is due to $\mu([o_1^k + \epsilon_1, o_2^k + \epsilon_0]) \leq \mu([0, 1]) = l$. Observe that

$$\mu\left([o_1^k + \epsilon_1, o_2^k + \epsilon_0]\right) - \mu\left(o_1^k\right) - 2\mu\left((o_1^k, o_1^k + \epsilon_1)\right) \geq \mu\left([o_1^k + \epsilon_1, o_2^k]\right) - \mu\left(o_1^k\right) - 2\mu\left((o_1^k, o_1^k + \epsilon_1)\right)$$
$$= \mu\left((o_1^k, o_2^k]\right) - \mu\left((o_1^k, o_1^k + \epsilon_1)\right) - \mu\left(o_1^k\right) - 2\mu\left((o_1^k, o_1^k + \epsilon_1)\right) = \mu\left((o_1^k, o_2^k]\right) - \mu\left(o_1^k\right) - 3\mu\left((o_1^k, o_1^k + \epsilon_1)\right)$$
$$> \delta - 3\frac{\delta}{6} = \frac{\delta}{2}, \quad (7)$$

where the last inequality follows from the contradiction assumption and Equation (4). Therefore, by substituting Equation (7) into Equation (6), we get

$$u_1(x_1, s_2) \leq \frac{l}{2k} - \frac{\epsilon_1}{2} \underbrace{\left( \mu\left([o_1^k + \epsilon_1, o_2^k + \epsilon_0]\right) - \mu\left(o_1^k\right) - 2\mu\left((o_1^k, o_1^k + \epsilon_1)\right) \right)}_{> \frac{\delta}{2}} + \frac{l\epsilon_0}{2}$$
$$< \frac{l}{2k} - \frac{\delta\epsilon_1}{4} + \frac{l\epsilon_0}{2} \stackrel{\text{Eq. (4)}}{<} \frac{l}{2k} - \frac{\delta\epsilon_1}{4} + \frac{l}{2}\frac{\delta\epsilon_1}{2l} = \frac{l}{2k}.$$

Hence we obtain a contradiction to Equation (3), as player 2 has a beneficial deviation. We therefore conclude that $\mu(o_1^k) \geq \mu((o_1^k, o_2^k])$. By symmetry we also conclude that $\mu(o_k^k) \geq \mu([o_{k-1}^k, o_k^k))$.

*Step 2:* We argue that for all $1 \leq i \leq k-2$, $\mu\left((o_i^k, o_{i+1}^k]\right) \geq \mu\left((o_{i+1}^k, o_{i+2}^k]\right)$.[4] Assume by contradiction that $\mu\left((o_{i+1}^k, o_{i+2}^k]\right) - \mu\left((o_i^k, o_{i+1}^k]\right) = \delta > 0$ holds for some $i$. Consider

$$s_2 = \left(o_1^k + \epsilon_0, \ldots, o_i^k + \epsilon_0, o_{i+1}^k + \epsilon_1, o_{i+2}^k + \epsilon_0, \ldots, o_k^k + \epsilon_0\right)$$

---
[4] Step 2 is needed only if $k \geq 3$.

for $\epsilon_1, \epsilon_0 > 0$ such that $\mu(o_i^k + \epsilon_j) = 0$ for all $i \in [k]$, $j \in \{0, 1\}$ and

$$\mu\left((o_{i+1}^k, o_{i+1}^k + \epsilon_1)\right) < \frac{\delta}{4}, \quad \epsilon_0 < \frac{\delta \epsilon_1}{4l}. \tag{8}$$

Namely, $s_2$ is similar to $o^k$, but every facility is shifted right by $\epsilon_0$, except from the facility on $o_{i+1}^k$ which is shifted right by $\epsilon_1$. See Figure 1 for illustration. With a construction similar to the one given in Equation (5), we have

$$u_1(s_1, s_2) \leq \frac{l}{2k} + \sum_{f \in [0, o_1^k) \cap s_1} \frac{\epsilon_0}{2} + \sum_{f \in [o_1^k, o_1^k + \epsilon_1) \cap s_1} \epsilon_0 + \sum_{f \in (o_i^k + \epsilon_0, o_{i+1}^k + \epsilon_1) \cap s_1} \frac{\epsilon_1 - \epsilon_0}{2} + \sum_{f \in (o_{i+1}^k + \epsilon_1, o_{i+2}^k + \epsilon_0) \cap s_1} \frac{\epsilon_0 - \epsilon_1}{2}. \tag{9}$$

Taking expectation over Equation (9), we get

$$u_1(x_1, s_2) \leq \frac{l}{2k} + \frac{\epsilon_0}{2}\mu\left([0, o_1^k)\right) + \epsilon_0 \mu\left([o_1^k, o_1^k + \epsilon_0)\right) + \frac{\epsilon_1 - \epsilon_0}{2}\mu\left((o_i^k + \epsilon_0, o_{i+1}^k + \epsilon_1)\right) + \frac{\epsilon_0 - \epsilon_1}{2}\mu\left((o_{i+1}^k + \epsilon_1, o_{i+2}^k + \epsilon_0)\right).$$

Lemma 4 guarantees that $\mu\left([0, o_1^k)\right) = 0$; thus

$$\begin{aligned}
u_1(x_1, s_2) &\leq \frac{l}{2k} + \epsilon_0 \mu\left([o_1^k, o_1^k + \epsilon_0)\right) + \frac{\epsilon_1 - \epsilon_0}{2}\mu\left((o_i^k + \epsilon_0, o_{i+1}^k + \epsilon_1)\right) \\
&\quad + \frac{\epsilon_0 - \epsilon_1}{2}\mu\left((o_{i+1}^k + \epsilon_1, o_{i+2}^k + \epsilon_0)\right) \\
&\leq \frac{l}{2k} - \frac{\epsilon_1}{2}\left(\mu\left((o_{i+1}^k + \epsilon_1, o_{i+2}^k + \epsilon_0)\right) - \mu\left((o_i^k + \epsilon_0, o_{i+1}^k + \epsilon_1)\right)\right) + l\epsilon_0,
\end{aligned} \tag{10}$$

where the last inequality is due to

$$\epsilon_0 \left(\mu\left([o_1^k, o_1^k + \epsilon_0)\right) - \frac{1}{2}\mu\left((o_i^k + \epsilon_0, o_{i+1}^k + \epsilon_1)\right) + \frac{1}{2}\mu\left((o_{i+1}^k + \epsilon_1, o_{i+2}^k + \epsilon_0)\right)\right) \leq \mu([0, 1])\epsilon_0 = l\epsilon_0.$$

Observe that

$$\begin{aligned}
&\mu\left((o_{i+1}^k + \epsilon_1, o_{i+2}^k + \epsilon_0)\right) - \mu\left((o_i^k + \epsilon_0, o_{i+1}^k + \epsilon_1)\right) \geq \mu\left((o_{i+1}^k + \epsilon_1, o_{i+2}^k)\right) - \mu\left((o_i^k, o_{i+1}^k + \epsilon_1)\right) \\
&= \mu\left((o_{i+1}^k, o_{i+2}^k]\right) - \mu\left((o_{i+1}^k, o_{i+1}^k + \epsilon_1)\right) - \mu\left((o_i^k, o_{i+1}^k]\right) - \mu\left((o_{i+1}^k, o_{i+1}^k + \epsilon_1)\right) \\
&= \mu\left((o_{i+1}^k, o_{i+2}^k]\right) - \mu\left((o_i^k, o_{i+1}^k]\right) - 2\mu\left((o_{i+1}^k, o_{i+1}^k + \epsilon_1)\right) \\
&> \delta - 2\frac{\delta}{4} = \frac{\delta}{2},
\end{aligned} \tag{11}$$

where the last inequality follows from the contradiction assumption and Equation (8). By substituting Equation (11) into Equation (10), we get

$$u_1(x_1, s_2) \leq \frac{l}{2k} - \frac{\epsilon_1}{2}\underbrace{\left(\mu\left((o_{i+1}^k + \epsilon_1, o_{i+2}^k + \epsilon_0)\right) - \mu\left((o_i^k + \epsilon_0, o_{i+1}^k + \epsilon_1)\right)\right)}_{>\frac{\delta}{2}} + l\epsilon_0$$

$$< \frac{l}{2k} - \frac{\delta\epsilon_1}{4} + l\epsilon_0 \stackrel{\text{Eq. (8)}}{<} \frac{l}{2k} - \frac{\delta\epsilon_1}{4} + l\frac{\delta\epsilon_1}{4l} = \frac{l}{2k},$$

Hence player 2 has a beneficial deviation, and we obtain a contradiction. Therefore, for all $1 \leq i \leq k-2$ we have $\mu\left((o_i^k, o_{i+1}^k]\right) \geq \mu\left((o_{i+1}^k, o_{i+2}^k]\right)$. By symmetry, we further conclude that

$$\forall 1 \leq i \leq k-2: \ \mu\left([o_i^k, o_{i+1}^k)\right) \leq \mu\left([o_{i+1}^k, o_{i+2}^k)\right).$$

*Step 3:* From Steps 1,2 and monotonicity of $\mu$, we have

- $\mu\left(o_1^k\right) \geq \mu\left((o_1^k, o_2^k]\right) \geq \mu\left((o_2^k, o_3^k]\right) ..... \geq \mu\left((o_{k-2}^k, o_{k-1}^k]\right) \geq \mu\left((o_{k-1}^k, o_k^k]\right) \geq \mu\left(o_k^k\right)$
- $\mu\left(o_1^k\right) \leq \mu\left([o_1^k, o_2^k)\right) \leq \mu\left([o_2^k, o_3^k)\right) ..... \leq \mu\left([o_{k-2}^k, o_{k-1}^k)\right) \leq \mu\left([o_{k-1}^k, o_k^k)\right) \leq \mu\left(o_k^k\right)$

Using the above inequalities we conclude that $\mu\left(o_i^k\right) = \frac{1}{k}$ for all $i \in [k]$, i.e. $x_1$ is $(l,k)$-SOI; hence C1 is satisfied. From Corollary 2 we know that in case $x_2 \neq o^k$, $u_2(x_1, x_2) < 1 - \frac{l}{2k}$; hence C2 is also satisfied. □

Corollary 3 follows immediately from Theorem 2.

COROLLARY 3. *In any equilibrium profile, facilities are materialized in $o^k$ with probability 1.*

**5. $N$ Players** In this section, we analyze $N$-player games with $N > 2$. We show that some multi-unit games possess a pure Nash equilibrium, unlike (asymmetric) two-player games. First, we show necessary and sufficient conditions for a profile to be in equilibrium, which provides full characterization of the structure of equilibrium in case it exists. Then, we characterize the multi-unit games which possess a pure equilibrium. Finally, we identify mixed strategy profiles for some games that lack pure strategy equilibrium.

**5.1. Notations for this section** We hereby refer to pure location Hotelling games as *single-unit games*, in the sense that for all $i \in [N]$ $n_i = 1$, i.e. each player has a single facility. For convenience, we denote $n \stackrel{\text{def}}{=} \sum_{i=1}^N n_i$. Also recall that $\mathcal{L}(\mathbf{s}) = \cup_{i \in [N]} s_i$. For brevity, we often refer to a facility located in $f \in [0,1]$ simply as facility $f$.

Given a pure strategy profile $\mathbf{s} = (s_1, \ldots, s_N)$, facilities located in $\min(\mathcal{L}(\mathbf{s}))$ or in $\max(\mathcal{L}(\mathbf{s}))$ are referred to as *peripheral facilities*. A facility $f$ owned by player $i$ is called a *lone facility* if player $i$ is the only player that locates a facility in $f$ under $\mathbf{s}$, i.e. $|\{j \in [N] : j \neq i, f \in s_j\}| = 0$. Alternatively, if $|\{j \in [N] : j \neq i, f \in s_j\}| = 1$, $f$ is called a *paired facility*.

Facilities $f_1, f_2 \in \mathcal{L}(\mathbf{s})$ are called *neighbors* if $f_1 \neq f_2$ and $(f_1, f_2) \cap \mathcal{L}(\mathbf{s}) = \emptyset$.[5] $f_l \in \mathcal{L}(\mathbf{s})$ is called a *left neighbor* of $f \in \mathcal{L}(\mathbf{s})$ if $f_l < f$ and $(f_l, f) \cap \mathcal{L}(\mathbf{s}) = \emptyset$. In other words, $f_l \in \mathcal{L}(\mathbf{s})$ is a left neighbor of $f \in \mathcal{L}(\mathbf{s})$ if $f_l$ is among the closest facilities to $f$ from its left side. Analogously, $f_r \in \mathcal{L}(\mathbf{s})$ is a right neighbor of $f \in \mathcal{L}(\mathbf{s})$ if $f < f_r$ and $(f, f_r) \cap \mathcal{L}(\mathbf{s}) = \emptyset$. Note that there can be more than one left/right neighbor to every $f \in \mathcal{L}(\mathbf{s})$, and there may be none.

We denote by $c_r(f; \mathbf{s})$ the quantity of customers traveling left in order to reach a facility (or facilities) located in $f \in [0,1]$. Namely,

$$c_r(f; \mathbf{s}) = \begin{cases} \int_f^1 \mathbb{1}_{d(t,f) \leq d(t, \mathcal{L}(\mathbf{s}))} dt & f \in \mathcal{L}(\mathbf{s}) \\ 0 & otherwise \end{cases}.$$

Similarly, $c_l(f; \mathbf{s})$ denotes the quantity of customers traveling right in order to reach a facility (or facilities) located in $f \in [0,1]$,

$$c_l(f; \mathbf{s}) = \begin{cases} \int_0^f \mathbb{1}_{d(t,f) \leq d(t, \mathcal{L}(\mathbf{s}))} dt & f \in \mathcal{L}(\mathbf{s}) \\ 0 & otherwise \end{cases}.$$

When analyzing a multi-unit game, we consider the following associated single-unit game:

DEFINITION 4 (FLATTENED GAME, FLATTENED PROFILE). Given a multi-unit game $G = \left(N, (n_i)_{i \in [N]}\right)$ and a pure strategy profile $\mathbf{s} \in \prod_{i=1}^N S_i$,

- the *flattened game of $G$*, denoted $\tilde{G}$, is a single-unit game with $n = \sum_{i=1}^N n_i$ players. That is, $\tilde{G} = \left(n, (1)_{i \in [n]}\right)$;
- a *flattened profile of $\mathbf{s}$* is a pure strategy profile $\tilde{\mathbf{s}} = (\tilde{s}_1, \ldots, \tilde{s}_n) \in [0,1]^n$ in the corresponding flattened game $\tilde{G}$, where for every $i \in [N], j \in [n_i]$ there exists a unique $k \in [n]$ such that $\tilde{s}_k = s_i^j$.[6]

---

[5] Note that if $f_1, f_2$ are paired with each other, $f_1$ is not a neighbor of $f_2$ and vice versa.

[6] Formally, there is a bijection $M : \{(i,j) : i \in [N], j \in [n_i]\} \to [n]$ such that $s_i^j = \tilde{s}_{M(i,j)}$.

We hereinafter use the tilde notations when referring to objects of flattened games and single-unit games, while $\mathbf{s}$ and $G$ refer to general multi-unit games. Note that $\tilde{G}$ is a symmetric game, and several flattened profiles exist for every strategy of $\mathbf{s}$, all of which are identical up to renaming the players.

For example, let $G = (2,(2,1))$ and let $\mathbf{s} = \left(\{\frac{1}{4}, \frac{3}{4}\}, \{\frac{1}{2}\}\right)$. It follows that $\tilde{G} = (3,(1,1,1))$, and two flattened profiles of $\mathbf{s}$ are $\left(\frac{1}{4}, \frac{3}{4}, \frac{1}{2}\right)$ and $\left(\frac{3}{4}, \frac{1}{4}, \frac{1}{2}\right)$.

**5.2. Pure strategy equilibrium characterization** The following result summarizes what is known about pure equilibria in single-unit games (see, e.g., [7]).

THEOREM 3. *Let $\tilde{G} = \left(N, (1)_{i \in [N]}\right)$. Necessary and sufficient conditions for a pure strategy profile $\tilde{\mathbf{s}}$ to be in equilibrium are*
1. *The two peripheral (leftmost/rightmost) facilities are paired.*
2. *For all $i \in [N]$ it holds that $u_i(\tilde{\mathbf{s}}) \geq \max_{f \in \mathcal{L}(\tilde{\mathbf{s}}), \sigma \in \{l,r\}} c_\sigma(f; \tilde{\mathbf{s}})$.*

In this subsection we identify necessary and sufficient conditions for a strategy profile to be in equilibrium in a multi-unit game. Clearly, the conditions of Theorem 3 must be somehow taken into account, since single-unit games are a special case of multi-unit games. We begin by showing that the first condition of Theorem 3 is vital regardless of the number of facilities players can locate.

PROPOSITION 4. *If $\mathbf{s}$ is in equilibrium in $G$, then the peripheral facilities are paired.*

PROOF. W.l.o.g. let the leftmost facility $f_l$ be a lone facility of player $i \in [N]$, such that $f_l = s_i^1 \in s_i$. Let $f' \in \min \mathcal{L}(\mathbf{s}_{-i})$ be the leftmost facility of players other than $i$, and denote by $s_i^j$ the left neighbor of $f'$, which belongs to player $i$ (possibly $j = 1$). For $s_i' \in S_i$ such that

$$s_i' = s_i \setminus \{s_i^j\} \cup \left\{\frac{s_i^j + f'}{2}\right\},$$

it follows that $u_i(\mathbf{s}) < u_i(s_i', \mathbf{s}_{-i})$; hence we obtain a contradiction to $\mathbf{s}$ being in equilibrium. □

Note that the second condition of Theorem 3 implies that if $\tilde{\mathbf{s}}$ is a pure equilibrium of $\tilde{G}$, then for every paired facility $f \in \mathcal{L}(\mathbf{s})$ it holds that $c_l(f; \mathbf{s}) = c_r(f; \mathbf{s})$. The following preposition shows the same applies in multi-unit games.

PROPOSITION 5. *If $\mathbf{s}$ is in equilibrium in $G$, then for every paired facility $f \in \mathcal{L}(\mathbf{s})$ it holds that $c_l(f; \mathbf{s}) = c_r(f; \mathbf{s})$.*

PROOF. Let $f \in s_i$ be a paired facility of player $i$, and w.l.o.g. let $c_l(f; \mathbf{s}) - c_r(f; \mathbf{s}) = \epsilon > 0$. Notice that $\mathcal{V}_i(f; \mathbf{s}) = \frac{c_l(f;\mathbf{s}) + c_r(f;\mathbf{s})}{2}$. Consider

$$s_i' = s_i \setminus \{f\} \cup \{f - \epsilon\}.$$

If $f$ has a left neighbor of player $i$ under $\mathbf{s}$, it follows that

$$u_i(s_i', \mathbf{s}_{-i}) = u_i(\mathbf{s}) - \mathcal{V}_i(f; \mathbf{s}) + c_l(f; \mathbf{s}) - \frac{\epsilon}{2} = u_i(\mathbf{s}) + \frac{1}{2} \underbrace{(c_l(f; \mathbf{s}) - c_r(f; \mathbf{s}) - \epsilon)}_{> 0} > u_i(\mathbf{s}).$$

Alternatively, if $f$ has no left neighbor of player $i$ under $\mathbf{s}$, $u_i(s_i', \mathbf{s}_{-i})$ is even greater. In both cases player $i$ has a beneficial deviation; hence $\mathbf{s}$ is not in equilibrium. □

The lemmas below will make use of the above propositions. While the previous two propositions extended the conditions of Theorem 3 to multi-unit games, Lemma 5 analyzes a scenario which can only occur if player(s) can locate more than one facility.

LEMMA 5. *If $\mathbf{s}$ is in equilibrium in $G$, then no lone facility has a neighbor owned by the same player.*

PROOF. First, assume that player $i$ has two (or more) neighboring lone facilities. Denote by $s_i^l, s_i^r$ the leftmost and rightmost lone facilities in a chain of neighboring lone facilities owned by player $i$ (possibly $r = l+1$, i.e. the length of the chain is two). Proposition 4 implies that $s_i^l, s_i^r$ are not peripheral. Let $(f_l, f_r)$ be the smallest interval such that $f_l, f_r \in \mathcal{L}(\mathbf{s}_{-i})$ and $s_i^l, s_i^r \in (f_l, f_r)$. Observe that

$$s_i' = s_i \setminus \{s_i^l, s_i^r\} \cup \left\{ \frac{f_l + s_i^l}{2}, \frac{s_i^r + f_r}{2} \right\}$$

is a beneficial deviation for player $i$, which cannot occur if $\mathbf{s}$ is an equilibrium profile. This is true even if player $i$ owns facilities in $f_l$ and/or $f_r$. Thus, there are no neighboring lone facilities owned by the same player.

Alternatively, denote by $s_i^j$ a lone facility owned by player $i$, and let $s_i^{j+1}$ be its right paired neighbor (symmetric for left neighbor). Here again, Proposition 4 suggests that $s_i^j$ is not peripheral. There are two possible sub-cases:
1. If the left neighbor(s) of $s_i^j$ is (are) not owned by player $i$, a beneficial deviation can be constructed by shifting the facility in $s_i^j$ to the left.
2. If $s_i^{j-1} \in s_i$ is a left neighbor of $s_i^j$, then $s_i^{j-1} \in s_i$ must be paired (according to the previous case). Let

$$s_i' = s_i \setminus \{s_i^{j-1}, s_i^j, s_i^{j+1}\} \cup \left\{\epsilon, s_i^{j-1} + \frac{\epsilon}{2}, s_i^{j+1} - \frac{\epsilon}{2}\right\}$$

for $\epsilon$ such that $0 < \epsilon < f_l = \min(\mathcal{L}(\mathbf{s}))$. Due to Proposition 5, it holds that $c_l(s_i^{j-1}; \mathbf{s}) = c_r(s_i^{j-1}; \mathbf{s})$ and $c_l(s_i^{j+1}; \mathbf{s}) = c_r(s_i^{j+1}; \mathbf{s})$. Therefore,

$$\mathcal{V}_i\left(s_i^{j-1} + \frac{\epsilon}{2}; s_i', \mathbf{s}_{-i}\right) + \mathcal{V}_i\left(s_i^{j+1} - \frac{\epsilon}{2}; s_i', \mathbf{s}_{-i}\right) = c_r(s_i^{j-1}; \mathbf{s}) + c_l(s_i^{j+1}; \mathbf{s}) - \frac{\epsilon}{2}$$
$$\stackrel{\text{Prop.5}}{=} \mathcal{V}_i(s_i^{j-1}; \mathbf{s}) + \mathcal{V}_i(s_i^j; \mathbf{s}) + \mathcal{V}_i(s_i^{j+1}; \mathbf{s}) - \frac{\epsilon}{2}. \quad (12)$$

Moreover, notice that

$$\mathcal{V}_i(\epsilon; s_i', \mathbf{s}_{-i}) + \mathcal{V}_i\left(s_i^1; s_i', \mathbf{s}_{-i}\right) \geq \mathcal{V}_i(\epsilon; s_i', \mathbf{s}_{-i}) + \mathcal{V}_i(s_i^1; \mathbf{s}) - \frac{f_l + \epsilon}{4} = \mathcal{V}_i(s_i^1; \mathbf{s}) + \frac{f_l + \epsilon}{4}$$
$$\stackrel{f_l > \epsilon}{>} \mathcal{V}_i(s_i^1; \mathbf{s}) + \frac{\epsilon}{2}. \quad (13)$$

By combining Equations (12) and (13), we get

$$\mathcal{V}_i\left(s_i^{j-1} + \frac{\epsilon}{2}; s_i', \mathbf{s}_{-i}\right) + \mathcal{V}_i\left(s_i^{j+1} - \frac{\epsilon}{2}; s_i', \mathbf{s}_{-i}\right) + \mathcal{V}_i(\epsilon; s_i', \mathbf{s}_{-i}) + \mathcal{V}_i\left(s_i^1; s_i', \mathbf{s}_{-i}\right)$$
$$> \mathcal{V}_i(s_i^{j-1}; \mathbf{s}) + \mathcal{V}_i(s_i^j; \mathbf{s}) + \mathcal{V}_i(s_i^{j+1}; \mathbf{s}) + \mathcal{V}_i(s_i^1; \mathbf{s}).$$

Finally, $\mathcal{V}_i(f; s_i', \mathbf{s}_{-i}) = \mathcal{V}_i(f; \mathbf{s})$ for every $f \in s_i \setminus \{s_i^1, s_i^{j-1}, s_i^j, s_i^{j+1}\}$; thus

$$u_i(s_i', \mathbf{s}_{-i}) = \sum_{f \in s_i'} \mathcal{V}_i(f; s_i', \mathbf{s}_{-i}) > \sum_{f \in s_i} \mathcal{V}_i(f; \mathbf{s}) = u_i(\mathbf{s}),$$

and $s_i'$ is a beneficial deviation for player $i$.
Overall, we showed that in equilibrium a lone facility cannot have a neighbor owned by the same player. $\square$

At this point, we link multi-unit games and their auxiliary single-unit games. We claim that a profile $\mathbf{s}$ can be in equilibrium in $G$ only if $\tilde{\mathbf{s}}$ is an equilibrium profile of $\tilde{G}$.

LEMMA 6. *If $\mathbf{s}$ is in equilibrium in $G$, then $\tilde{\mathbf{s}}$ is in equilibrium in $\tilde{G}$.*

PROOF. Assume by contraposition that $\tilde{\mathbf{s}}$ is not in equilibrium in $\tilde{G}$, so at least one of the conditions of Theorem 3 does not hold. By Proposition 4 the peripheral facilities of $\mathbf{s}$ are paired; hence by definition in the flattened profile the peripheral facilities of $\tilde{\mathbf{s}}$ are paired as well.

Therefore, there exists $i' \in [n]$ such that

$$q = u_{i'}(\tilde{\mathbf{s}}) < \max_{\sigma \in \{l,r\}, f' \in \mathcal{L}(\tilde{\mathbf{s}})} c_\sigma(f'; \tilde{\mathbf{s}}) = p.$$

Let $i \in [N], j \in [n_i]$ such that $\tilde{s}_{i'} = s_i^j$, and denote by $(l, f)$ a maximum point of $c_\sigma(f'; \tilde{\mathbf{s}})$, i.e. $c_l(f; \tilde{\mathbf{s}}) = p$ (similar analysis applies if $c_\sigma(f'; \tilde{\mathbf{s}})$ attains its maximum only for $\sigma = r$). In addition, observe that $u_{i'}(\tilde{\mathbf{s}}) = \mathcal{V}_i(s_i^j; \mathbf{s})$. There are three possible cases:

1. If $f \notin s_i$, for

$$s_i' = s_i \setminus \{s_i^j\} \cup \{f - \epsilon\}$$

   where $\epsilon > 0$ is a small enough constant we obtain $u_i(s_i', \mathbf{s}_{-i}) > u_i(\mathbf{s})$.

2. $f \in s_i$ is a lone facility owned by player $i$. It follows from Proposition 4 and Lemma 5 that $f$ is not peripheral and has no neighbor owned by player $i$. Hence, $f$ has right and left neighbors, $f_r, f_l$ respectively, such that

$$\mathcal{V}_i(f; \mathbf{s}) = \frac{f_r - f_l}{2} = p.$$

   Let $s_i' \in S_i$ such that

$$s_i' = s_i \setminus \{s_i^j, f\} \cup \{f_l + \epsilon, f_r - \epsilon\}$$

   for $\epsilon > 0$, and notice that for every $f' \in s_i \cap s_i'$ it holds that $\mathcal{V}_i(f'; s_i', \mathbf{s}_{-i}) \geq \mathcal{V}_i(f'; \mathbf{s})$. Therefore,

$$u_i(s_i', \mathbf{s}_{-i}) \geq u_i(\mathbf{s}) - (q + p) + 2p - \epsilon > u_i(\mathbf{s})$$

   for small enough $\epsilon$.

3. $f \in s_i$ is paired. By Proposition 5 $c_l(f; \mathbf{s}) = c_r(f; \mathbf{s}) = p/2$. Let $s_i' \in S_i$ such that

$$s_i' = s_i \setminus \{s_i^j, f\} \cup \{f - \epsilon, f + \epsilon\},$$

   for $\epsilon > 0$. We shall analyze the worst case scenario for player $i$, when $f$'s neighbors belong also to player $i$. It holds that

$$u_i(s_i', \mathbf{s}_{-i}) \geq u_i(\mathbf{s}) - \epsilon - (q + p) + 2p - \epsilon > u_i(\mathbf{s}).$$

In all cases player $i$ has a beneficial deviation; therefore we obtain contradiction since $\mathbf{s}$ is not an equilibrium in $G$. □

The following lemma addresses another scenario which may occur only in multi-unit games.

LEMMA 7. *If $\mathbf{s}$ is in equilibrium in $G$, then all the facilities owned by the same player attract the same customer mass.*[7] *Formally,*

$$\forall i \in [N], \forall f, f' \in s_i : \quad \mathcal{V}_i(f; \mathbf{s}) = \mathcal{V}_i(f'; \mathbf{s}). \tag{14}$$

PROOF. Assume by contraposition that $\mathbf{s}$ is an equilibrium profile, and the condition set in Equation (14) does not hold. Therefore, there exists a player $i$ whose facilities are located in $f, f' \in s_i$ such that

$$\mathcal{V}_i(f; \mathbf{s}) - \mathcal{V}_i(f'; \mathbf{s}) = \epsilon > 0.$$

---

[7] Facilities owned by different players, however, can attract different customer mass.

If the facility located in $f$ is lone under $\mathbf{s}$, it follows from Proposition 4 that it is not peripheral, and from Lemma 5 that its neighbors, denoted $f_l$ and $f_r$, do not belong to player $i$. For $s_i' = s_i \setminus \{f, f'\} \cup \{f_l + \frac{\epsilon}{2}, f_r - \frac{\epsilon}{2}\}$, it holds that

$$u_i(s_i', \mathbf{s}_{-i}) = u_i(\mathbf{s}) - \mathcal{V}_i(f; \mathbf{s}) - \mathcal{V}_i(f'; \mathbf{s}) + \mathcal{V}_i\left(f_l + \frac{\epsilon}{2}; s_i', \mathbf{s}_{-i}\right) + \mathcal{V}_i\left(f_r - \frac{\epsilon}{2}; s_i', \mathbf{s}_{-i}\right)$$
$$= u_i(\mathbf{s}) - \mathcal{V}_i(f; \mathbf{s}) - \mathcal{V}_i(f'; \mathbf{s}) + 2\mathcal{V}_i(f; \mathbf{s}) - \frac{\epsilon}{2}$$
$$= u_i + \frac{\epsilon}{2}.$$

Hence $s_i'$ is a beneficial deviation of player $i$.

Alternatively, if the facility located in $f$ is paired, it does not have lone neighbors belonging to player $i$ under $\mathbf{s}$. Further, from Proposition 5 it follows that $c_l(f; \mathbf{s}) = c_r(f; \mathbf{s}) = \mathcal{V}_i(f; \mathbf{s})$. Denote $s_i' = s_i \setminus \{f, f'\} \cup \{f - \delta, f + \delta\}$ for $\delta > 0$. Observe that

$$\mathcal{V}_i(f - \delta; s_i', \mathbf{s}_{-i}) + \mathcal{V}_i(f + \delta; s_i', \mathbf{s}_{-i}) = c_l(f; \mathbf{s}) + c_r(f; \mathbf{s}) - \delta.$$

In addition,

$$\sum_{f'' \in s_i' \cap s_i} \mathcal{V}_i(f''; s_i', \mathbf{s}_{-i}) \geq \sum_{f'' \in s_i' \cap s_i} \mathcal{V}_i(f''; \mathbf{s}) - \frac{\delta}{2},$$

with equality in the case that the facility located in $f$ does not have any neighbors belonging to player $i$ under $\mathbf{s}$. Overall,

$$u_i(s_i', \mathbf{s}_{-i}) = \sum_{f'' \in s_i'} \mathcal{V}_i(f''; s_i', \mathbf{s}_{-i}) \geq \sum_{f'' \in s_i' \cap s_i} \mathcal{V}_i(f''; \mathbf{s}) - \frac{\delta}{2} + c_l(f; \mathbf{s}) + c_r(f; \mathbf{s}) - \delta$$
$$= \sum_{f'' \in s_i' \cap s_i} \mathcal{V}_i(f''; \mathbf{s}) + 2\mathcal{V}_i(f; \mathbf{s}) - \frac{3\delta}{2} = \sum_{f'' \in s_i} \mathcal{V}_i(f''; \mathbf{s}) + \epsilon - \frac{3\delta}{2} = u_i(\mathbf{s}) + \epsilon - \frac{3\delta}{2}.$$

For $\delta < \frac{2\epsilon}{3}$, we conclude that $s_i'$ is a beneficial deviation of player $i$. Hence, we obtained a contradiction to $\mathbf{s}$ being in equilibrium.
□

We deduce from Lemmas 5,6,7 three necessary conditions that must hold in every equilibrium profile. The following lemma shows that these conditions are not only necessary, but also sufficient.

LEMMA 8. *If under a strategy profile* $\mathbf{s}$
- *no lone facility has a neighbor of the same player;*
- *all the facilities owned by the same player attract the same customer mass; and*
- *$\tilde{\mathbf{s}}$ is in equilibrium in $\tilde{G}$,*

*then* $\mathbf{s}$ *is in equilibrium in $G$.*

PROOF. Assume by contradiction that $\mathbf{s}$ is not in equilibrium in $G$, and player $i$ has a beneficial deviation. Observe that

$$\sum_{f \in s_i \cap s_i'} \mathcal{V}_i(f; \mathbf{s}) + \sum_{f \in s_i \setminus s_i'} \mathcal{V}_i(f; \mathbf{s}) = u_i(\mathbf{s}) < u_i(s_i', \mathbf{s}_{-i}) = \sum_{f \in s_i' \cap s_i} \mathcal{V}_i(f; s_i', \mathbf{s}_{-i}) + \sum_{f' \in s_i' \setminus s_i} \mathcal{V}_i(f'; s_i', \mathbf{s}_{-i}). \tag{15}$$

For every $f \in s_i \cap s_i'$, it holds that $\mathcal{V}_i(f; s_i', \mathbf{s}_{-i}) \leq \mathcal{V}_i(f; \mathbf{s})$, since $f$ has no lone neighbors of player $i$ under $\mathbf{s}$. Summing over all the elements of $s_i \cap s_i'$, we get

$$\sum_{f \in s_i \cap s_i'} \mathcal{V}_i(f; s_i', \mathbf{s}_{-i}) \leq \sum_{f \in s_i \cap s_i'} \mathcal{V}_i(f; \mathbf{s}). \tag{16}$$

From Equations (15) and (16) it follows that

$$\sum_{f \in s_i \setminus s'_i} \mathcal{V}_i(f; \mathbf{s}) < \sum_{f' \in s'_i \setminus s_i} \mathcal{V}_i(f'; s'_i, \mathbf{s}_{-i}).$$

Notice that $|s'_i \setminus s_i| = |s_i \setminus s'_i|$. Therefore, there exist $f \in s_i \setminus s'_i, f' \in s'_i \setminus s_i$ such that

$$\mathcal{V}_i(f; \mathbf{s}) < \mathcal{V}_i(f'; s'_i, \mathbf{s}_{-i}). \tag{17}$$

Let $i'$ be an index of a player in $\tilde{G}$ such that $\tilde{s}_{i'} = f$. Recall that $\tilde{\mathbf{s}} = (\tilde{s}_1, \ldots \tilde{s}_n)$ is in equilibrium in $\tilde{G}$; hence $f'$ is not a beneficial deviation for player $i'$, i.e. $\mathcal{V}_{i'}(f'; f', \tilde{\mathbf{s}}_{-i'}) \leq \mathcal{V}_{i'}(f; \tilde{\mathbf{s}})$. Thus,

$$\mathcal{V}_i(f'; s_i \setminus \{f\} \cup \{f'\}, \mathbf{s}_{-i}) \leq \mathcal{V}_i(f; \mathbf{s}). \tag{18}$$

Equations (17),(18) imply that

$$\mathcal{V}_i(f'; s_i \setminus \{f\} \cup \{f'\}, \mathbf{s}_{-i}) < \mathcal{V}_i(f'; s'_i, \mathbf{s}_{-i}). \tag{19}$$

We proceed by showing that Equation (19) cannot hold, which will contradict the assumption of player $i$ having a beneficial deviation. Let $(f_l, f_r)$ be the smallest interval such that $f' \in (f_l, f_r)$ and $f_l, f_r \in \mathcal{L}(\mathbf{s}_{-i}) \cup \{0, 1\}$. Notice that $f_l, f_r$ do not depend on $s_i$. Since player $i$ does not have two neighboring lone facilities under $\mathbf{s}$, we know that $|s_i \cap (f_l, f_r)| \leq 1$. In addition, $|s_i \cap (f_l, f_r)| = 0$ cannot hold, since otherwise we would have $\mathcal{V}_i(f'; s_i \setminus \{f\} \cup \{f'\}, \mathbf{s}_{-i}) = \mathcal{V}_i(f'; s'_i, \mathbf{s}_{-i})$, which contradicts Equation (19).

Therefore, $|s_i \cap (f_l, f_r)| = 1$. Denote this facility $f^* \in s_i \cap (f_l, f_r)$, and w.l.o.g. assume $f' < f^*$ ($f' > f$ and $f' = f$ are analyzed similarly). The means of choosing $f_l$ and $f_r$ ensure that player $i$'s facility in $f^*$ is lone under $s_i$. Now,

- If $f^* \in s'_i$, it follows that

$$\mathcal{V}_i(f'; s_i \setminus \{f\} \cup \{f'\}, \mathbf{s}_{-i}) = \frac{f^* - f_l}{2} \geq \mathcal{V}_i(f'; s'_i, \mathbf{s}_{-i}),$$

which contradicts Equation (19).

- Otherwise, if $f^* \notin s'_i$, it holds that

$$\mathcal{V}_i(f'; s'_i, \mathbf{s}_{-i}) \leq \frac{f_r - f_l}{2} = \mathcal{V}_i(f^*; \mathbf{s}), \tag{20}$$

with equality if $f'$ is the only facility owned by player $i$ under $s'_i$ inside $(f_l, f_r)$. From the second condition of this lemma we know that $\mathcal{V}_i(f; \mathbf{s}) = \mathcal{V}_i(f^*; \mathbf{s})$. By combining this fact with Equations (17) and (20), we get

$$\mathcal{V}_i(f; \mathbf{s}) < \mathcal{V}_i(f'; s'_i, \mathbf{s}_{-i}) \leq \frac{f_r - f_l}{2} = \mathcal{V}_i(f^*; \mathbf{s}) = \mathcal{V}_i(f; \mathbf{s}),$$

which yields a contradiction.

We conclude that players do not have beneficial deviations under $\mathbf{s}$, indicating that $\mathbf{s}$ is an equilibrium profile. □

By gathering the results obtained in Lemmas 5-8, one concludes that

THEOREM 4. *In a multi-unit game G, necessary and sufficient conditions for a pure strategy profile $\mathbf{s}$ to be in equilibrium are:*
1. *No lone facility has a neighboring facility of the same player under $\mathbf{s}$.*
2. *All the facilities owned by the same player attract the same customer mass.*

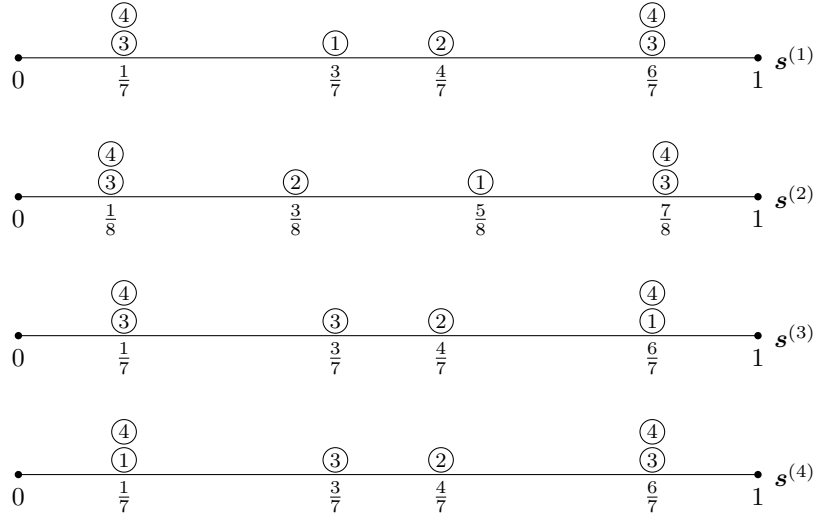

FIGURE 2. Four pure strategy profiles for $G = (4, (1, 1, 2, 2))$. Circles represent the chosen locations of the players; for example player 1 has a lone facility in $\frac{3}{7}$ in profile $\mathbf{s}^{(1)}$, lone facility in $\frac{5}{8}$ in profile $\mathbf{s}^{(2)}$, paired facility in $\frac{6}{7}$ in profile $\mathbf{s}^{(3)}$ and paired facility in $\frac{1}{7}$ in $\mathbf{s}^{(4)}$. Observe that flattened profiles of all the above are in equilibrium in the flattened game $\tilde{G}$. Profiles $\mathbf{s}^{(1)}$ and $\mathbf{s}^{(2)}$ are in equilibrium, while $\mathbf{s}^{(3)}$ and $\mathbf{s}^{(4)}$ are not. First, notice that the conditions of Lemma 8 are met under $\mathbf{s}^{(1)}$ and $\mathbf{s}^{(2)}$. On the other hand, $\mathbf{s}^{(3)}$ is not in equilibrium since the lone facility of player 3 in $\frac{3}{7}$ has a neighbor of hers in $\frac{1}{7}$. While $u_3(\mathbf{s}^{(3)}) = \frac{5}{14}$, in case she deviates to $s_3' = \left(\frac{1}{7} + \epsilon, \frac{4}{7} - \epsilon\right)$ her payoff is $u_3(s_3', \mathbf{s}_{-3}^{(3)}) = \frac{3}{7} - \epsilon$. Under $\mathbf{s}^{(4)}$, it holds that $\frac{1}{7} = \mathcal{V}_3(\frac{6}{7}; \mathbf{s}^{(4)}) < \mathcal{V}_3(\frac{3}{7}; \mathbf{s}^{(4)}) = \frac{1.5}{7}$; thus condition 2 of Lemma 8 is not satisfied. Here too $s_3' = \left(\frac{1}{7} + \epsilon, \frac{4}{7} - \epsilon\right)$ is a beneficial deviation for player 3.

3. $\tilde{\mathbf{s}}$ *is in equilibrium in* $\tilde{G}$.

PROOF. One direction follows from Lemmas 5, 6 and 7, while the other direction follows from Lemma 8. □

Figure 2 illustrates two equilibrium profiles ($\mathbf{s}^{(1)}$ and $\mathbf{s}^{(2)}$) and two profiles which are not in equilibrium for the game $G = (4, (1, 1, 2, 2))$. Note that infinite equilibria exist for this game, and can be obtained not only by swapping facilities between the players (while preserving the necessary and sufficient conditions), but also by modifying the set of locations.

**5.3. Pure strategy equilibrium existence** Until now, we focused on determining if a profile is an equilibrium for a specific game, and only eluded determining whether a game possesses a pure equilibrium or not. In this subsection, we classify the set of games possessing pure equilibria. Recall that our two-player analysis showed that if one of the players owns more facilities than the other, there is no pure equilibrium. The following notion extends this idea of dominance to multiple players. We remind the reader that we assume w.l.o.g. $n_1 \leq n_2 \leq \cdots \leq n_N$, and that the total number of facilities is denoted by $n$, i.e. $n = \sum_{i=1}^{N} n_i$.

DEFINITION 5 (DOMINANT PLAYER). In a multi-unit game, player $N$ is called a *dominant player* if she owns more than half of all facilities, i.e.

$$\sum_{i=1}^{N-1} n_i < n_N.$$

Before we deal with any general $n$, note that possible multi-unit games with $n \leq 3$ are:
1. $G = (2, (1, 1))$, which is a single-unit game with two players. As studied in [9], the only equilibrium of this game is obtained for the pure strategy profile $\left(\frac{1}{2}, \frac{1}{2}\right)$.

2. $G = (2,(1,2))$, which is an instance of a two-player game with $l = n_1 = 1, k = n_2 = 2$. As indicated in the previous section, this game has no pure equilibrium.
3. $G = (3,(1,1,1))$, namely a single-unit game with three players. This is known to have no pure equilibrium (see [5]), but mixed equilibrium does exist (as studied in [16]).

Leaving these games aside, it turns out that knowledge on the existence of a dominant player is the only property necessary for determining the existence or absence of a pure equilibrium in a multi-unit game. The following result is in accordance with the results of the previous section, which implied that if $n_1 < n_2$ in a two-player game, there is no pure equilibrium.

THEOREM 5. *In a multi-unit game $G$ with $n \geq 4$, there exists a pure Nash equilibrium if and only if it has no dominant player.*

The proof of Theorem 5 relies on several supporting claims. Proposition 6 shows that if a dominant player exists, there is no pure equilibrium.

PROPOSITION 6. *Let $G$ be a multi-unit game with $n \geq 4$. If player $N$ is a dominant player, then there is no pure equilibrium.*

PROOF. Assume by contradiction that player $N$ is a dominant player and $\mathbf{s}$ is a pure strategy equilibrium of $G$. Since $n_N > n - n_N$, there are facilities owned by player $N$ which are lone. Let $p$ denote the number of lone facilities player $N$ owns under $\mathbf{s}$. According to Theorem 4, a lone facility owned by player $N$ cannot have a neighbor which also belongs to her, and cannot be peripheral. Thus,

$$|\{f : i \in [N-1], f \in s_i, f \text{ has a lone neighbor of player } i\}| \geq p+1.$$

In addition, player $N$ has $n_N - p$ paired facilities. Therefore

$$|\{f : i \in [N-1], f \in s_i, f \text{ is paired with } f' \in s_N\}| = n_N - p.$$

Overall, we conclude that the number of facilities is at least

$$p + (p+1) + 2(n_N - p) = 2n_N + 1 > n,$$

which yields the desired contradiction. □

Proving the other direction of Theorem 5, namely that there exists at least one pure strategy equilibrium for every game with a dominant player, is of a different flavor and deserves a different treatment. We divide the class of all multi-unit games into two sub-classes: even and odd $n$, for reasons that will become apparent. We first deal with the case of an even number of facilities.

LEMMA 9. *Let $G$ be a multi-unit game with $n \geq 4$ where player $N$ is not dominant. If $n$ is even, then there exists a pure equilibrium.*

PROOF. Denote by $(f_k)_{k=1}^n$ the sequence

$$\forall k \in [n]: \quad f_k = \frac{2(k \pmod{\frac{n}{2}}) - 1}{n}.$$

Notice that $f_k = f_{k+\frac{n}{2}}$ for every $k \leq \frac{n}{2}$, and if $i \leq \frac{n}{2}$, $i \neq k$ then $f_i \neq f_k$. Therefore, the sequence contains $\frac{n}{2}$ pairs of identical elements, and each pair is different than the other pairs. Now, we allocate $f_1, \ldots f_{n_1}$ to player 1, $f_{n_1+1}, \ldots, f_{n_1+n_2}$ to player 2 and so on. Namely,

$$s_i = \left(f_{\sum_{k=1}^{i-1} n_k + 1}, f_{\sum_{k=1}^{i-1} n_k + 2}, \ldots, f_{\sum_{k=1}^{i} n_k}\right).$$

Denote $\mathbf{s} = (s_1, \ldots, s_N)$. For every $k \leq \frac{n}{2}$, the sub-sequence $(f_k, f_{k+1}, \ldots, f_{k+\frac{n}{2}-1}, f_{k+\frac{n}{2}})$ contains $\frac{n}{2} + 1$ elements; therefore, $f_k$ and $f_{k+\frac{n}{2}}$ are owned by different players, since there is no dominant player. Every facility attracts the same customer mass under $\mathbf{s}$, players do not have lone facilities and $\tilde{\mathbf{s}}$ is in equilibrium in $\tilde{G}$; therefore, Theorem 4 implies that $\mathbf{s}$ is in equilibrium. □

Constructing a pure equilibrium for games with an even $n$ is straightforward, as indicated by Lemma 9 above. However, if $n$ is odd, there cannot be a profile where all facilities are paired; thus, there must be at least one lone facility. Recall that every equilibrium profile should also be in equilibrium in the corresponding flattened game. In Proposition 7 below we identify two constraints on the equilibrium structure.

PROPOSITION 7. *Let $G$ be a multi-unit game. If a profile $\tilde{\mathbf{s}}$ is in equilibrium in $\tilde{G}$, then*
1. *Players who own paired facilities obtain the same payoff.*
2. *If a player owns a lone facility with paired neighbors, she gets a strictly higher payoff than players who own paired facilities.*

PROOF. To prove the first claim, recall that if player $i$ (in the flattened game) owns a paired facility in $f$, then $u_i(\tilde{\mathbf{s}}) = \frac{c_r(f;\tilde{\mathbf{s}}) + c_l(f;\tilde{\mathbf{s}})}{2}$. In addition, $c_r(f;\tilde{\mathbf{s}}) = c_l(f;\tilde{\mathbf{s}})$ holds in equilibrium, since otherwise player $i$ has a beneficial deviation. Now, due to the second condition of Theorem 3, for every facility $f'$ owned by player $i'$ it holds that

$$u_i(\tilde{\mathbf{s}}) \geq \max_{f'' \in \mathcal{L}(\tilde{\mathbf{s}}), \sigma \in \{l,r\}} c_\sigma(f'';\tilde{\mathbf{s}}) \geq c_r(f';\tilde{\mathbf{s}}) = u_{i'}(\tilde{\mathbf{s}}), \quad u_{i'}(\tilde{\mathbf{s}}) \geq \max_{f'' \in \mathcal{L}(\tilde{\mathbf{s}}), \sigma \in \{l,r\}} c_\sigma(f'';\tilde{\mathbf{s}}) \geq c_r(f;\tilde{\mathbf{s}}) = u_i(\tilde{\mathbf{s}}).$$

Thus, $u_i(\tilde{\mathbf{s}}) = u_{i'}(\tilde{\mathbf{s}})$.

We move to the second claim. Let $f, f'$ be a lone facility and one of its neighboring paired facilities, owned by players $i$ and $i'$ respectively. Assume w.l.o.g. that $f < f'$. Since $\tilde{\mathbf{s}}$ is in equilibrium, $c_r(f';\tilde{\mathbf{s}}) = c_l(f';\tilde{\mathbf{s}})$; thus $u_{i'}(\tilde{\mathbf{s}}) = c_l(f';\tilde{\mathbf{s}})$. Due to symmetry, $c_r(f;\tilde{\mathbf{s}}) = c_l(f';\tilde{\mathbf{s}})$. Since $f$ is lone it is not peripheral; thus $c_l(f;\tilde{\mathbf{s}}) > 0$, and we have $u_i(\tilde{\mathbf{s}}) > c_r(f;\tilde{\mathbf{s}}) = u_{i'}(\tilde{\mathbf{s}})$.  □

Proposition 7 establishes two important constraints on the structure of equilibrium profiles in multi-unit games. First, assume that a player owns a lone facility with paired neighbors. Notice that all of her facilities must be lone as well: this is true since all the paired facilities attract the same customer mass, which is strictly less than that lone facility. Second, assume that a player owns both lone and paired facilities. If a lone facility of hers has paired neighbors (of other players), it obtains a strictly higher customer mass than what they obtain. Therefore, her lone facilities cannot have paired neighbors.

We shall use these two constraints in the construction of equilibrium profiles for odd $n$. While relatively simple equilibria exist for specific game instances, we are interested in a one-fits-all equilibrium structure. We are now ready to construct an equilibrium profile for multi-unit games with odd $n$.

LEMMA 10. *Let $G$ be a multi-unit game with $n \geq 5$ where player $N$ is not dominant. If $n$ is odd, then there exists a pure equilibrium.*

PROOF. We prove the lemma by constructing a pure equilibrium profile $\mathbf{s}$ for $G$, where all the facilities owned by player 1 are lone. The construction is done in steps, where in each step we sequentially locate a subset of all the facilities, and then allocate them to players. We will also present a visual snapshot of the $[0,1]$ segment after each step, where ① represents a facility allocated to player 1, and empty circles represent facilities of players $2, \ldots N-1$. Finally, we show that the profile $\mathbf{s}$ we constructed indeed satisfies the conditions of Theorem 4, and hence in equilibrium. Denote $p = \frac{1}{n+1}$.

*Step 1:* we locate the left peripheral facilities in $p$. We allocate these facilities to players $N$ and $N-1$. The current snapshot is

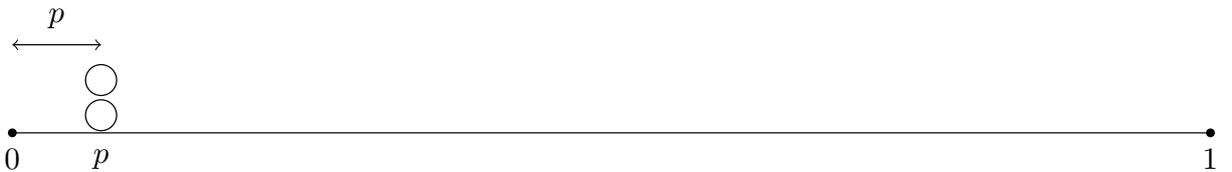

*Step 2:* We define the sequence $(f_i^{lone})_{i=1}^{2n_1-1}$ such that

$$f_i^{lone} = \begin{cases} (i+2)p & \text{if } i \text{ is odd} \\ (i+1)p + \frac{ip}{n_1} & \text{if } i \text{ is even} \end{cases}.$$

We allocate $(f_i^{lone})_{i=1}^{2n_1-1}$ in the following manner:
- For $i = 1$ to $2n_1 - 1$:
  — If $i$ is odd, assign $f_i^{lone}$ to player 1.
  — Otherwise, define the number of player $j$'s remaining facilities to be $n_j$ minus those allocated to her in Step 1 and iterations $1, \ldots i-1$. Assign $f_i^{lone}$ to the player with the highest number of remaining facilities, breaking ties by selecting the player with the lowest index.

Notice that the customer mass attracted to each even-indexed facility is $p$. The current snapshot is

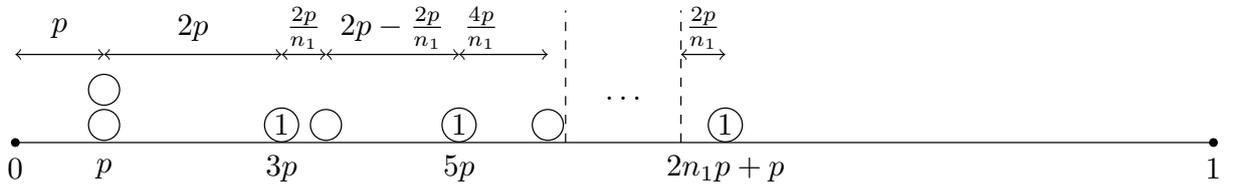

*Step 3:* So far we located and allocated $(2n_1 - 1) + 2$ facilities, which is an odd number. Thus, since $n$ is odd, there is an even number of facilities we did not locate yet. Denote by $(0, n_2', \ldots n_N')$ the number of remaining facilities of each player after the previous step.

CLAIM 1. *There is no dominant player in $(n_2', \ldots n_N')$.*

The proof contains a case analysis, and is deferred to the appendix. We define the sequence $(f_i^{pair})_{i=1}^{n-(2n_1+1)}$ such that

$$f_i^{pair} = 2n_1 p + 3p + 2p \left\lfloor \frac{i-1}{2} \right\rfloor.$$

Using a similar argument to the one given in Lemma 9, we can allocate $(f_i^{pair})_{i=1}^{n-(2n_1+1)}$ such that each pair contains facilities of different players. Notice that the rightmost facility is located in

$$f_{n-(2n_1+1)}^{pair} = 2n_1 p + 3p + 2p \left\lfloor \frac{n-2n_1-2}{2} \right\rfloor = 2n_1 p + 3p + p(n - 2n_1 - 3) = np = \frac{n}{n+1}.$$

The current snapshot is

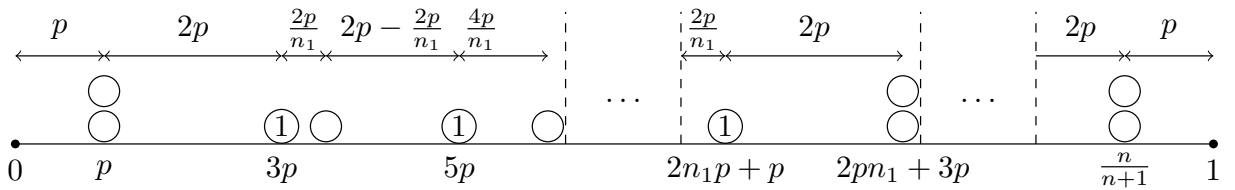

*Step 4:* We have located all $n$ facilities: two in Step 1; $2n_1 - 1$ in Step 2; and the remaining $n - (2n_1 + 1)$ in Step 3. We also allocated these facilities to the players such that each player $i$ owns $n_i$ facilities; thus **s** is a valid profile of $G$. We now show that the conditions of Theorem 4 hold for **s**:

1. Due to Step 2, every lone facility is either owned by player 1 and has neighbors of players $2, \ldots N$, or owned by a player $2, \ldots N$ and has neighboring facilities of player 1 only.

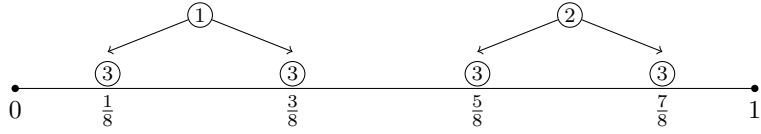

FIGURE 3. Mixed equilibrium for the game $(4,(1,1,4))$. Player 3, a dominant player, selects $o^4 = \left\{\frac{1}{8}, \frac{3}{8}, \frac{5}{8}, \frac{7}{8}\right\}$ deterministically. Player 1 selects $\frac{1}{8}$ with probability half, and $\frac{3}{8}$ with probability half. Player 3 selects $\frac{5}{8}$ with probability half, and $\frac{5}{8}$ with probability half. The payoff of players 1 and 2 is $\frac{1}{8}$ each, while player 3 gets $\frac{3}{4}$. Note the resemblance to the quasi-unique equilibrium profile of the two-player game $(2,(2,4))$.

2. Due to Steps 1 and 3, all paired facilities obtain a customer mass of $p$. In addition, the construction in Step 2 assures that
   - for every odd index $i, 1 \leq i < 2n_1 - 1$, it holds that $f_{i+2}^{lone} - f_i^{lone} = 2p$;
   - for every even index $i, 2 \leq i < 2n_1$, it holds that $f_{i+2}^{lone} - f_i^{lone} = 2p + \frac{2p}{n_1}$; and
   - $c_l(f_1^{lone}; \mathbf{s}) + c_r(f_1^{lone}; \mathbf{s}) = c_l(f_{2n_1-1}^{lone}; \mathbf{s}) + c_r(f_{2n_1-1}^{lone}; \mathbf{s}) = 2p + \frac{2p}{n_1}$.
   
   Therefore, every lone facility of players $2, \ldots N$ attracts a customer mass of exactly $p$ as well (recall that these were even-indexed facilities in the sequence $(f_i^{lone})_{i=1}^{2n_1-1}$, and each lies in a segment of length $2p$, from which it gets half). In addition, facilities owned by player 1 get $p(1 + \frac{1}{n_1})$ each (each lies in a segment of length $2p + \frac{2p}{n_1}$, from which it gets half). Hence, for every $i \in [N]$, player $i$'s facilities all attract the same customer mass.

3. In Step 3 we located (and allocated) at least one pair of facilities: since $N \geq 3$ and $n \geq 5$ we have $n \geq 3n_1 > 2n_1 + 1$. Hence, Steps 1 and 3 ensure that the left and right peripheral facilities are paired. Moreover, every player in the flattened game obtains a customer mass of at least $p$, and
$$p = \max_{f \in \mathcal{L}(\tilde{\mathbf{s}}), \sigma \in \{l,r\}} c_\sigma(f; \tilde{\mathbf{s}});$$
hence no player in the flattened game has a beneficial deviation. It follows from Theorem 3 that $\tilde{\mathbf{s}}$ is in equilibrium in $\tilde{G}$.

Overall, $\mathbf{s}$ is an equilibrium profile for $G$. □

To conclude this subsection, notice that Theorem 5 follows immediately from Proposition 6 and Lemmas 9 and 10.

**5.4. Mixed equilibria** We now show that the principles we relied upon in the previous sections for the analysis of two-player games may lead to constructing mixed equilibrium for some games with a dominant player. The key ingredient is that the strategies of players $1, 2, \ldots, N-1$ together mimic a $\left(\sum_{i=1}^{N-1} n_i, n_N\right)$-SOI strategy, as will be formally described next.

LEMMA 11. *Let $G = \left(N, (n_i)_{i \in [N]}\right)$ be a multi-unit game with a dominant player. If there exist constants $b_1, b_2, \ldots, b_{N-1} \in \mathbb{N}_{>0}$ such that*
- $\sum_{i=1}^{N-1} b_i = n_N$; *and*
- *for all $i \in [N-1]$ it holds that $\frac{n_i}{b_i} = \frac{n - n_N}{n_N}$*

*then $G$ possesses a mixed equilibrium.*

PROOF. We shall describe a mixed strategy profile $\mathbf{x}$, and show that players do not have profitable deviations. Let $o^{n_N} = \{o_1^{n_N}, o_2^{n_N} \ldots o_{n_N}^{n_N}\}$ as defined in Section 3. Since $\sum_{i=1}^{N-1} b_i = n_N$, there exists a partition of $o^{n_N}$, $B_1, B_2, \ldots, B_{N-1}$, such that $|B_i| = b_i$. For $i \in [N-1]$, let $x_i \in \Delta(S_i)$ such that
$$\mathbb{P}_{x_i}(s) = \begin{cases} 1/\binom{b_i}{n_i} & \text{if } s \subseteq B_i, |s| = n_i \\ 0 & \text{otherwise} \end{cases}.$$

We argue that $\mathbf{x} = (x_1, \ldots, x_{N-1}, o^{n_N})$ is in equilibrium. First, observe that for every $\mathbf{s}_{-N} \in supp(\mathbf{x}_{-N})$, it holds that $\mathcal{L}(\mathbf{s}_{-N}) \subset o^{n_N}$. Hence, for every $i \in [N-1]$, $\mathbf{s}_{-i} \in supp(\mathbf{x}_{-i})$, $s_i \in S_i$ and $f \in s_i$ it follows that

$$\mathcal{V}_i(f; s_i, \mathbf{s}_{-i}) \leq \frac{1}{2n_N}.$$

This implies for all $s_i \in S_i$, $u_i(s_i, \mathbf{x}_{-i}) \leq \frac{n_i}{2n_N} = u_i(\mathbf{x})$. We are therefore left to show that player $N$ has no beneficial deviation. Assume by contradiction that there exists $s'_N \in S_N$ such that $u_N(s'_N, \mathbf{x}_{-N}) > u_N(\mathbf{x}) = 1 - \frac{\sum_{i=1}^{N-1} n_i}{2n_N}$. Define an auxiliary two-player game

$$G' = \left(2, \left(\sum_{i=1}^{N-1} n_i, n_N\right)\right),$$

with players $1'$ and $2'$. Notice that $S_N$ (in $G$) is equal to $S_{2'}$ (in $G'$), and that $\mathbf{x}_{-N} \in \Delta(S_{1'})$. By the way we constructed $x_i$ for $i \in [N-1]$, it holds that

$$\mathbb{P}_{s_i \sim x_i}\left(o_i^j \in s_i\right) = \begin{cases} \frac{n_i}{b_i} & o_i^j \in B_i \\ 0 & otherwise \end{cases}.$$

Since $\frac{n_i}{b_i} = \frac{n - n_N}{n_N}$ and $(B_i)_{i=1}^{N-1}$ is a partition of $o^{n_N}$, it follows that

$$\forall j \in [n_N]: \quad \mathbb{P}_{\mathbf{s}_{-N} \sim \mathbf{x}_{-N}}\left(o_j^{n_N} \in \mathcal{L}(\mathbf{s}_{-N})\right) = \frac{n - n_N}{n_N} = \frac{\sum_{i=1}^{N-1} n_i}{n_N}.$$

Hence $\mathbf{x}_{-N}$ is a $\left(\sum_{i=1}^{N-1} n_i, n_N\right)$-SOI strategy of player $1'$ in $G'$. Recall that Corollary 2 implies that

$$\forall s_N \in S_N: \quad u_{2'}(\mathbf{x}_{-N}, s_N) = \sum_{\mathbf{s}_{-N} \in supp(\mathbf{x}_{-N})} \sum_{f \in s_N} \mathcal{V}_{2'}(f; \mathbf{s}_{-N}, s_N) \leq 1 - \frac{\sum_{i=1}^{N-1} n_i}{2n_N}, \tag{21}$$

while the contradiction assumption implies

$$1 - \frac{\sum_{i=1}^{N-1} n_i}{2n_N} < u_N(\mathbf{x}_{-N}, s'_N) = \sum_{\mathbf{s}_{-N} \in supp(\mathbf{x}_{-N})} \sum_{f \in s'_N} \mathcal{V}_{2'}(f; \mathbf{s}_{-N}, s'_N). \tag{22}$$

By using Equations (21) and (22) we get

$$1 - \frac{\sum_{i=1}^{N-1} n_i}{2n_N} < \sum_{\mathbf{s}_{-N} \in supp(\mathbf{x}_{-N})} \sum_{f \in s'_N} \mathcal{V}_{2'}(f; \mathbf{s}_{-N}, s'_N) \leq 1 - \frac{\sum_{i=1}^{N-1} n_i}{2n_N};$$

hence a contradiction is obtained. Therefore, player $N$ has no beneficial deviation in $\mathbf{x}$, thereby concluding the proof of Lemma 11. □

Figure 3 visualizes a mixed equilibrium for $G = (4, (1, 1, 4))$. As described in the Introduction, [10, Proposition 6] indicates the existence of mixed equilibria for the class of games $\{G = (N, (n_i)_{i \in [N]})\}$ for which $n_i = 1$ for $i \in [N-1]$, and there exists $k \in \mathbb{N}_{>1}$ such that $n_N = k(n - n_N)$. Indeed, such games are a special case of those characterized in Lemma 11, where $b_i = k$ for all $i \in [N-1]$.

In the proof of Lemma 11 we used the fact that $\mathbf{x}_{-N} = \prod_{i \in [N-1]} x_i$ was a $\left(\sum_{i=1}^{N-1} n_i, n_N\right)$-SOI strategy. Clearly, one can come up with strategies for players $1, 2, \ldots N-1$ such that the joint strategy $\mathbf{x}_{-N}$ will be a $\left(\sum_{i=1}^{N-1} n_i, n_N\right)$-SOI strategy, even without the lemma's conditions, by setting $x_i$ to be a $(n_i, n_N)$-SOI strategy. However, note that this is not enough, since unless the conditions of the lemma are met, with positive probability two "weak" players will position a facility in the same location. Thus, with positive probability, three facilities will be located at the same point, and so these players have beneficial deviations. Such a profile cannot be in equilibrium.

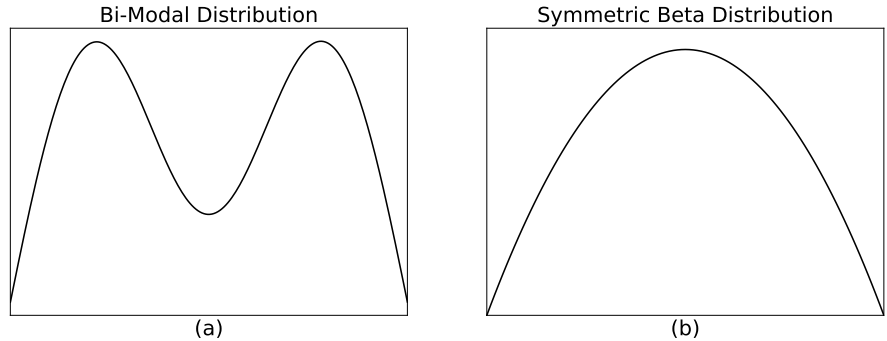

FIGURE 4. Examples of non-uniform distributions. In (a), the suggested profile will be in equilibrium, as both of the players have no profitable deviations. In (b), however, player 1 can deviate to $q_{1/2}$ and increase her payoff.

**6. Discussion and Future Work** We have presented a natural extension of pure location Hotelling games. For the case of two players, we showed the existence and quasi-uniqueness of equilibrium, where the strong player acts according to the socially optimal facility locations, and the weak player is mixing between subsets of the strong player's locations. Afterwards, for multi-player games, we extended the necessary and sufficient conditions required for a pure strategy profile to be in equilibrium in Hotelling games to the multi-unit setting. We characterized the set of games possessing pure equilibrium, and identified mixed equilibria for a large class of games which do not possess a pure equilibrium. The existence of equilibrium profiles for *any* such game, even under uniform distribution, is left as an open question.

Another direction to investigate is what happens when the customer distribution $g$ is not uniform, as studied in [13] for the single-unit Hotelling game.

Consider a two-player game with $n_1 = 1, n_2 = 2$. Let $q_a$ be the $a$-th quantile of $g$, namely

$$q_a = \inf\left\{ t^* \in (-\infty, \infty) : a \leq \int_{-\infty}^{t^*} g(t) dt \right\}.$$

A simple generalization of the equilibrium profile introduced in Section 3 is $(x_1, x_2)$, where

$$x_1 = \begin{cases} q_{1/4} & w.p. \frac{1}{2} \\ q_{3/4} & w.p. \frac{1}{2} \end{cases}, \qquad x_2 = (q_{1/4}, q_{3/4}).$$

If $g$ is Bi-Modal, it can be easily shown that the profile above is in fact in equilibrium. However, in case $g$ is the Symmetric Beta Distribution, the suggested profile is not in equilibrium, since player 1 would prefer choosing $q_{1/2}$ rather than mixing between $q_{1/4}$ and $q_{3/4}$. See Figure 4 for a visualization of this. We leave the question of equilibrium existence and structure in multi-unit facility location games under non-uniform distribution open as well.

Beyond the classical motivation of these games indicated in the introduction, new motivating scenarios in areas related to Data Science, such as Machine Learning, Advertising and Information Retrieval are emerging. Consider the following illustrative scenario. The readers of an on-line newspaper can be modeled as points on the $[0, 1]$ segment, where the point $p \in [0, 1]$ captures a reader with a desire for proportion $p$ of news and $(1-p)$ of editorials in a newspaper. The decision of a publisher is on the mix between news and editorials he will choose in his newspaper in order to attract as many readers as possible. A major aspect of on-line newspapers is that one may maintain several versions. This brings us into the context of multi-unit facility location games. Interestingly, recent work in data science [15] has shown the applicability of this approach in modeling realistic competition among publishers.

**Acknowledgments** This project has received funding from the European Research Council (ERC) under the European Union's Horizon 2020 research and innovation programme (grant agreement n° 740435).

**Appendix A: Omitted Proofs:**

**A.1. Proof of Proposition 1** Let $A = \{a_1, \ldots, a_k\}$, and w.l.o.g. $a_1 \leq \cdots \leq a_k$. For simplicity, denote $a_0 = 0, a_{k+1} = 1$. For $i \in [k+1]$, we let

$$b_i = a_i - a_{i-1}.$$

Note that $\sum_{i=1}^{k+1} b_i = 1$. It follows that

$$\begin{aligned} \text{SC}(A) &= \int_0^1 \min_{a \in A} d(t, a) dt \\ &= \int_{a_0=0}^{a_1} (a_1 - t) dt + \sum_{i=2}^{k} \int_{a_{i-1}}^{a_i} \min\{t - a_{i-1}, a_i - t\} dt + \int_{a_k}^{a_{k+1}=1} (t - a_{i-1}) dt. \end{aligned} \quad (23)$$

Observe that $\int_{a_0}^{a_1} (a_1 - t)dt = \frac{b_1^2}{2}$, and $\int_{a_k}^{a_{k+1}} (t - a_{i-1})dt = \frac{b_{k+1}^2}{2}$. In addition, for $i \in \{2, \ldots, k\}$ it holds that
$$\int_{a_{i-1}}^{a_i} \min\{t - a_{i-1}, a_i - t\}dt = \frac{b_i^2}{4}.$$

Therefore, Equation (23) can be rewritten as
$$\text{SC}(A) = \frac{b_1^2}{2} + \frac{b_{k+1}^2}{2} + \sum_{i=2}^{k} \frac{b_i^2}{4}. \tag{24}$$

To find the minimum of SC(A), one must constrain $(b_i)_{i \in [k+1]}$ to satisfy $\sum_{i=1}^{k+1} b_i = 1$. To do so, we define a new function and incorporate Lagrange Multipliers (see e.g. [3]):
$$h(b_1, \ldots, b_{k+1}, \lambda) = \frac{b_1^2}{2} + \frac{b_{k+1}^2}{2} + \sum_{i=2}^{k} \frac{b_i^2}{4} + \lambda \left( \sum_{i=1}^{k+1} b_i - 1 \right).$$

By assigning zeros to the partial derivatives of $h$, we get its minimum point:
$$b_i = \begin{cases} \frac{1}{2k} & i \in \{1, k+1\} \\ \frac{1}{k} & i \in \{2, 3..k\} \end{cases}.$$

Finally, since $b_i = a_i - a_{i-1}$, we obtain $A = \left( \frac{1}{2k}, \ldots, \frac{2i-1}{2k}, \ldots, \frac{2k-1}{2k} \right)$. $\square$

**A.2. Proof of Proposition 3.** Fix $x \in \Delta(S_1)$. For every set $A \subseteq [0,1]$, let
$$A_i = [0,1]^{i-1} \times A \times [0,1]^{l-i}.$$

Thus, an equivalent formulation of $\mu_x$ is
$$\mu_x(A) = \int_{[0,1]^l} \sum_{i=1}^{l} \mathbb{1}_{A_i} dx.$$

In order to show that $\mu$ is a measure, we need to show that it satisfies non-negativity, null empty set and countable additivity:
1. Non-negativity: $\forall A \subseteq [0,1]$, it follows that $\mathbb{1}_{A_i} \geq 0$; thus $\mu_x(A) \geq 0$.
2. Null empty set: if $A = \emptyset$, it follows that $A_i = \emptyset$ for all $1 \leq i \leq l$. Since $\mathbb{1}_\emptyset \equiv 0$, $\mu_x(\emptyset) = 0$.
3. Countable additivity: for disjoint sets $(B^j)_{j=1}^\infty$,

$$\mu_x \left( \bigcup_{j=1}^{\infty} B^j \right) = \int_{[0,1]^l} \sum_{i=1}^{l} \mathbb{1}_{\bigcup_{j=1}^{\infty} B^j} dx = \int_{[0,1]^l} \sum_{i=1}^{l} \sum_{j=1}^{\infty} \mathbb{1}_{B_i^j} dx \stackrel{\text{Fubini's Thm.}}{=} \sum_{j=1}^{\infty} \int_{[0,1]^l} \sum_{i=1}^{l} \mathbb{1}_{B_i^j} dx$$
$$= \sum_{j=1}^{\infty} \mu_x(B^j).$$

$\square$

PROPOSITION 8. *For any strategy $x_1$ of player 1 the following holds:*
1. *For any $0 \leq a < b \leq 1$, there exists $\epsilon > 0$ such that $a + \epsilon < b$ and $\mu(a + \epsilon) = 0$.*
2. *For any $a \in [0,1)$ and $\delta > 0$, there exists $\epsilon > 0$ such that $\mu((a, a + \epsilon)) < \delta$.*

PROOF.

1. Assume by contradiction that for any $\epsilon$ such that $0 < \epsilon < b - a$ it holds that $\mu(a+\epsilon) > 0$. Observe that
$$\left[\frac{b-a}{4}, \frac{b-a}{2}\right] \subset (0, b-a)$$
is closed and $\mu$ is bounded, $\mu$ obtains a minimum point inside it,
$$c = \min_{y \in \left[\frac{b-a}{4}, \frac{b-a}{2}\right]} \mu(a+y).$$
The contradiction assumption implies that $c > 0$; thus
$$\mu\left(\left[\frac{b-a}{4}, \frac{b-a}{2}\right]\right) \geq \sum_{n=1}^{\infty} \mu\left(\frac{b-a}{4}\left(1+\frac{1}{n}\right)\right) \geq \sum_{n=1}^{\infty} c = \infty.$$
However, $\mu\left(\left[\frac{b-a}{4}, \frac{b-a}{2}\right]\right) \leq \mu([0,1]) = l$; hence a contradiction is obtained.

2. Define a decreasing sequence of sets $(B_i)_{i=1}^{\infty}$, $B_i \in \mathcal{B}([0,1])$ such that
$$B_i = \left(a, a + \frac{1}{n_0 + i}\right),$$
where $n_0$ is the minimal integer satisfying $a + \frac{1}{n_0} \leq 1$. Since $B_i$ decreases to $\emptyset$ and $\mu$ is a measure, we know that $\lim_{i \to \infty} \mu(B_i) = \mu(\emptyset) = 0$. Therefore, there exists $j$ such that
$$\mu(B_j) = \mu\left(a, a + \frac{1}{n_0 + j}\right) < \delta.$$
Hence we denote $\epsilon = \frac{1}{n_0 + j}$ and the claim holds.
□

### A.3. Proof of Lemma 4.

1. The assertion is a property of fixed-sum games. In particular, recall that Lemma 2 implies that $u_1(o^{l,k}, x_2) \geq \frac{l}{2k}$ for any $x_2 \in \Delta(S_2)$; thus in case player 1 gets less, she can deviate to $o^{l,k}$. A similar argument applies for player 2.

2. If $\mu_{x_1}([0, o_1^k)) > 0$, with positive probability a facility owned by player 1 is realized in $[0, o_1^k)$; hence
$$\mathbb{P}_{s_1 \sim x_1}\left(\exists f \in s_1 : \mathcal{V}_1(f; s_1, o^k) < \frac{1}{2k}\right) > 0.$$
Recall that Lemma 2 implies that $\mathcal{V}_1(f; s_1, o^k) \leq \frac{1}{2k}$ for every $s_1 \in S_1$ and every $f \in s_1$; thus together we have
$$u_1(x_1, o^k) = \mathbb{E}_{s_1 \sim x_1} u_1(s_1, o^k) = \mathbb{E}_{s_1 \sim x_1}\left(\sum_{f \in s_1} \mathcal{V}_1(f; s_1, o^k)\right) < \frac{l}{2k}.$$
Obviously, if $x_2 = o^k$ we obtain a contradiction to part one of this lemma. Alternatively, if $x_2 \neq o^k$ it follows that $(x_1, x_2)$ is not an equilibrium, since $u_2(x_1, o^k) > 1 - \frac{l}{2k}$ and therefore player 2 has a beneficial deviation. We conclude that $\mu_{x_1}([0, o_1^k)) = 0$, and by similar arguments we have $\mu_{x_1}((o_k^k, 1]) = 0$.

3. By Lemma 2, $u_1(x_1, o^k) \leq \frac{l}{2k}$, and when combined with part one of this lemma we get $u_1(x_1, o^k) \leq u_1(x_1, x_2)$. In addition, if $u_1(x_1, o^k) < u_1(x_1, x_2)$ then $u_2(x_1, o^k) > u_2(x_1, x_2)$; hence $(x_1, x_2)$ is not an equilibrium.
□

**A.4. Proof of Claim 1.** Let $j^*$ be the index of the player with the highest number of facilities (if two or more players own the highest number of facilities under $(n'_2, \ldots n'_N)$, there is no dominant player). We have two possible cases:

1. If $j^* < N$: since initially $n_{j^*} \leq n_N$ and the tie breaker is inclined towards low indices, it follows from the construction that $n'_{j^*} = n'_N + 1$. Since $n'_{j^*} + n'_N$ is odd while $\sum_{j=2}^{N} n'_j$ is even, there exist at least one more player, denoted by $i$, with $n'_i \geq 1$. Hence

$$n'_{j^*} = n'_N + 1 \leq n'_N + n'_i \leq \sum_{j=2, j \neq j^*}^{N} n'_j,$$

 and player $j^*$ is not a dominant player.

2. If $j^* = N$, we separate the analysis into two cases:

 (a) If during Step 2 we did not allocate facilities to players $2, \ldots, N-1$, then exactly $n_1$ facilities owned by player $N$ are allocated by the end of Step 2. In addition, $n_1 + 1$ facilities of players $2, \ldots N-1$ are allocated by the end of Step 2. Hence,

 $$n_N - n_1 < \sum_{j=1}^{N-1} n_j - n_1 \Rightarrow n'_N = n_N - n_1 \leq \sum_{j=1}^{N-1} n_j - n_1 - 1 = \sum_{j=2}^{N-1} n'_j.$$

 (b) Otherwise, let $i^*$ denote the latest iteration a facility was allocated to a player with index from $2, \ldots, N-1$, and denote by $k$ the index of this player. It follows that before iteration $i^*$ player $k$ had $n'_k + 1$ facilities. In addition, after iteration $i^*$ facilities were only allocated to player $N$; thus $n'_N \leq n'_k + 1$. Since $n'_N > n'_k$ (recall that $j^* = N$), we have $n'_N = n'_k + 1$. Notice that $n'_N + n'_k$ is odd while $\sum_{j=2}^{N} n'_j$ is even; thus there exists at least one more player, denoted by $i$, with $n'_i \geq 1$. Hence

 $$n'_N = n'_k + 1 \leq n'_k + n'_i \leq \sum_{j=2}^{N-1} n'_j.$$

Overall, we showed that there is no dominant player in $(n'_2, \ldots n'_N)$, which concludes the proof of this claim. □